\documentclass[12pt]{article}

\usepackage{amsmath}
\usepackage{amsthm}
\usepackage{amssymb}

\usepackage{mathrsfs}
\usepackage{fancyhdr}
\usepackage{pb-diagram} 

\usepackage{graphicx}
\usepackage{natbib}

\usepackage{tcolorbox} 
\tcbuselibrary{breakable, skins, theorems}
\usepackage[margin=1in]{geometry}
\usepackage{setspace}
\usepackage{sectsty}
\sectionfont{\centering}

\usepackage{amssymb,amsmath,amsfonts,eurosym,geometry,ulem,graphicx,caption,color,setspace,sectsty,comment,footmisc,natbib,pdflscape,array,bm,tabularx, mathtools,bbm,multirow}
\usepackage[colorlinks,citecolor=red,urlcolor=blue,bookmarks=false,hypertexnames=true]{hyperref}

\usepackage{threeparttable}
\usepackage{subfig}
\usepackage{booktabs}
\PassOptionsToPackage{hyphens}{url}
\usepackage[hyphens]{url}

\newtheorem{proposition}{Result}
\usepackage[at]{easylist}
\usepackage{}
\title{
Individual Rationality Conditions of Identifying Matching Costs in Transferable Utility Matching Games}

\author{Suguru Otani\thanks{suguru.otani@e.u-tokyo.ac.jp, Market Design Center, University of Tokyo\\Declarations of interest: none} }
\begin{document}
\maketitle

\begin{abstract}
The widely applied method for measuring assortativeness in a transferable utility matching game is the matching maximum score estimation proposed by Fox (2010). This article reveals that by combining unmatched agents, transfers, and individual rationality conditions with sufficiently large penalty terms, it's possible to identify the coefficient parameter of a single common constant, i.e., matching costs in the market.
\end{abstract} 

\section{Introduction}
One of the most well-known methods for measuring matching assortativeness was developed by \cite{fox2010qe,fox2018qe}. Although the method allows researchers to flexibly incorporate additional inequalities from equilibrium conditions, the additional information that helps identification is unknown. This article focuses on pairwise stability in a transferable utility (TU) matching model and investigates the identification by adding unmatched agents, transfer data, and individual rationality (IR) conditions, which are available in many empirical applications.
This article theoretically and numerically reveals that using unmatched agents, transfers, and individual rationality conditions with sufficiently large penalty terms makes it possible to identify the coefficient parameter of a single common constant, i.e.,  matching costs in the market. 

First, transfer data is known to improve identification power practically if it is available. The working paper version of \cite{fox2013aej} uses bidding price data in FCC Spectrum Auction in a single market setting. They numerically confirm that if the number of agents is small or the variance of errors is large, bidding price data generated from a tatonnement process drastically reduces the bias and RMSE (the root of mean squared error) of an estimated single parameter. \cite{akkus2015ms} use acquisition prices between buyer firms and target firms as transfer data. \cite{pan2017determinants} uses CEO compensations between CEOs and firms. Appendix A of \cite{akkus2015ms} shows Monte Carlo evidence that adding transfers not only improves the accuracy of estimation but also enables researchers to identify the coefficient of a non-interacted term.
In another strand, there is some theoretical and empirical evidence that using unmatched agents helps identification. \cite{fox2018jpe} show nonparametric identification results in two-sided matching model. Section 5 of their paper proves that the distribution of unobserved complementarities conditional on observed characteristics can be recovered if each firm on each side is part of exactly one match or is unmatched in each feasible assignment. \cite{otani2021coalitional} uses unmatched firms to identify and estimate matching costs in a one-sided one-to-many coalitional mergers in Japanese shipping industry with Monte Carlo simulations. 
This article shows how these data affect construction of inequalities for identification and estimation. 

Second, individual rationality (IR, henceforth) conditions, i.e., the binary choice information are necessary for identification of a fixed cost \citep{bresnahan1990entry}. In the TU matching game, IR conditions capture whether agents choose to stay matched or unmatched. This article shows under what conditions matching fixed costs can be identified and estimated.

\section{Model}\label{sec:baseline_model}

\subsection{Baseline matching model}
I consider a two-sided one-to-one TU merger matching game in a single market. Let $\mathcal{N}_b$ and $\mathcal{N}_s$ be the sets of potential finite buyers and sellers respectively. Let $b=1,\cdots,|\mathcal{N}_b|$ be buyer firms and let $s=1,\cdots,|\mathcal{N}_s|$ be seller firms where $|\cdot|$ is cardinality. Let $\mathcal{N}_{b}^{m}$ denote the set of ex-post matched buyers and $\mathcal{N}_{b}^{u}$ denote that of ex-post unmatched buyers such that $\mathcal{N}_b= \mathcal{N}_{b}^{m}\cup\mathcal{N}_{b}^{u}$ and $\mathcal{N}_{b}^{m}\cap\mathcal{N}_{b}^{u}=\emptyset$. For the seller side, define $\mathcal{N}_{s}^{u}$ and $\mathcal{N}_{s}^{m}$ as the set of ex-post matched and unmatched sellers such that $\mathcal{N}_s= \mathcal{N}_{s}^{m}\cup\mathcal{N}_{s}^{u}$ and $\mathcal{N}_{s}^{m}\cap\mathcal{N}_{s}^{u}=\emptyset$. Let $\mathcal{M}^m$ be the sets of all ex-post matched pairs $(b,s)\in\mathcal{N}_{b}^{m}\times \mathcal{N}_{s}^{m}$. Let $\mathcal{M}$ denote the set of all ex-post matched pairs $(b,s)\in\mathcal{M}^{m}$ and unmatched pairs $(\tilde{b},\emptyset)$ and $(\emptyset,\tilde{s})$ for all $\tilde{b}\in \mathcal{N}_b^u$ and $\tilde{s}\in \mathcal{N}_s^u$ where $\emptyset$ means a null agent generating unmatched payoff. 

Each firm can match at most one agent on the other side, so  $|\mathcal{N}_b^{m}|=|\mathcal{N}_s^{m}|$. The matching joint production function is defined as $f(b,s)=V_b(b,s)+V_s(b,s)$ where $V_b:\mathcal{M}\rightarrow \mathbb{R}$ and $V_s:\mathcal{M}\rightarrow \mathbb{R}$. The net matching values for buyer $b$ and seller $s$ are defined as $V_b(b,s)=f(b,s)-p_{b,s}$ and $V_s(b,s)+p_{b,s}$, where $p_{b,s}\in \mathbb{R}_{+}$ is the equilibrium merger price paid to seller firm $s$ by buyer firm $b$ and $p_{b\emptyset}=p_{\emptyset s}=0$. For scale normalization, I assume $V_b(b,\emptyset)=0$ and $V_s(\emptyset,s)=0$ for all $b\in \mathcal{N}_b$ and $s\in \mathcal{N}_s$. 

The matching allocation is incentive compatible given equilibrium merger prices if each agent maximizes its profit. 
The allocation is feasible if agents' excess demand for counterpart agents equals zero.
Under some technical assumptions, \cite{azevedo2018existence} show that the matching allocation with equilibrium prices is a competitive equilibrium if the allocation is incentive compatible given the equilibrium prices and is feasible.
The competitive equilibrium is equivalent to the stable matching.\footnote{See Proposition 3.3 of \cite{galichon2018optimal} for reference.}
At competitive equilibrium, each buyer maximizes $V_b(b,s)$ across seller firms, whereas each seller maximizes $V_s(b,s)$ across buyer firms. 

Specifically, the stability conditions for buyer firm $b \in \mathcal{N}_b$ and seller firm $s \in \mathcal{N}_s$ are as follows:
\begin{align}
    V_b(b,s) &\ge V_b(b,s') \quad \forall s' \in \mathcal{N}_s \cup \emptyset,s'\neq s,\label{eq:stability_ineq}\\
    V_s(b,s) &\ge V_s(b',s) \quad \forall b' \in \mathcal{N}_b\cup \emptyset,b'\neq b.\nonumber
\end{align}

Based on Equation \eqref{eq:stability_ineq} and equilibrium price conditions $p_{b',s}\le p_{b,s}$ and $p_{b,s'}\le p_{b',s'}$ in \cite{akkus2015ms}, I construct the inequalities for matches $(b,s)\in \mathcal{M}$ and $(b',s')\in \mathcal{M}, (b',s')\neq(b,s)$ as follows:
\begin{align}
    f(b,s)-f(b,s')&\ge p_{b,s}-p_{b,s'}\ge p_{b,s}-p_{b',s'},\label{eq:pairwise_stable_ineq}\\
    f(b',s')-f(b',s)&\ge p_{b',s'}-p_{b',s}\ge p_{b',s'}-p_{b,s},\nonumber\\
    V_s(b,s)-V_s(b',s)&\ge 0,\nonumber\\
    V_{s'}(b',s')-V_s(b,s')&\ge 0,\nonumber
\end{align}
where $p_{b',s}$ and $p_{b,s'}$ are unrealized equilibrium merger prices that cannot be observed in the data. The last two inequalities cannot be derived from the data because the researchers cannot observe how the total matching value $f(b,s)$ is shared between buyer $b$ and seller $s$.

\subsection{Transfer and unmatched data.}

In many empirical applications, researchers do not obtain data on transfers and unmatched firms. The minimum number of possible inequalities is discussed in Result \ref{prop:num_of_all_ineq}.
\begin{proposition}\label{prop:num_of_all_ineq}
    Suppose that researchers obtain (A) transfer data $p_{bs}$ for matched buyer firm $b$ and seller firm $s$ for all $b\in\mathcal{N}_b^{m}$ and $s\in\mathcal{N}_s^{m}$, and (B) unmatched data about $\mathcal{N}_b^{u}$ and $\mathcal{N}_s^{u}$ with their observed characteristics. Let $_nC_k$ be the number of combinations of $k$ objects from $n$ objects, and $_nP_k$ be the number of permutations, representing the different ordered arrangements of a $k$-element subset of an $n$-set.\footnote{For reference, $_nC_k=\frac{n!}{k!(n-k)!}$ and $_nP_k=\frac{n!}{(n-k)!}$ where $!$ is a factorial function.} Then, the followings hold:
    \begin{enumerate}
        \renewcommand{\labelenumi}{(\roman{enumi})}
        \item Researchers can construct the following inequalities for matches $(b,s)\in \mathcal{M}$ and $(b',s')\in \mathcal{M}, (b',s')\neq(b,s)$:
        \begin{align}
            f(b,s)-f(b,s')&\ge p_{b,s}-p_{b',s'}.\label{eq:inequality_with_transfer}
        \end{align}
        The minimum number of total inequalities is $_{|\mathcal{M}^m|+|\mathcal{N}_b^u|+|\mathcal{N}_s^u|}P_2$.
        \item If the data does not contain (A), researchers can construct the following inequalities for matches $(b,s)\in \mathcal{M}$ and $(b',s')\in \mathcal{M}, (b',s')\neq(b,s)$:
        \begin{align}
            f(b,s)+ f(b',s')&\ge f(b,s')+f(b',s).\label{eq:inequality_without_transfer}
        \end{align}
        The minimum number of total inequalities is $_{|\mathcal{M}^m|+|\mathcal{N}_b^u|+|\mathcal{N}_s^u|}C_2$.
        \item If the data does not contain (B), researchers can construct Inequality \eqref{eq:inequality_without_transfer} for matches $(b,s)\in \mathcal{M}^m$ and $(b',s')\in \mathcal{M}^m, (b',s')\neq(b,s)$. The minimum number of total inequalities is $_{|\mathcal{M}^m|}P_2$.
        \item If the data does not contain (A) and (B), researchers can construct Inequality \eqref{eq:inequality_with_transfer} for matches $(b,s)\in \mathcal{M}^m$ and $(b',s')\in \mathcal{M}^m, (b',s')\neq(b,s)$. The minimum number of total inequalities is $_{|\mathcal{M}^m|}C_2$.
    \end{enumerate}
\end{proposition}

\subsection{Individual rationality conditions}

IR conditions are implicitly included in Inequality \eqref{eq:inequality_with_transfer} when transfer data is available because $\mathcal{M}$ restores unmatched buyer $\tilde{b}$ and seller $\tilde{s}$ as $(\tilde{b},\emptyset)$ and $(\emptyset,\tilde{s})$, respectively. For matched pair $(b,s)\in \mathcal{M}$ and unmatched buyer $(\tilde{b},\emptyset)\in \mathcal{M}$, Inequality \eqref{eq:inequality_with_transfer} gives $f(b,s)-f(b,\emptyset)=f(b,s)\ge p_{b,s}-p_{\tilde{b},\emptyset}\ge p_{b,s}$, i.e., 
\begin{align}
    f(b,s)-p_{b,s}&\ge 0. \label{eq:IR_condition} 
\end{align}
Importantly, unlike pairwise inequalities, the IR condition holds for each matched firm on each side regardless of the availability of transfer data. To distinguish IR conditions from pairwise inequalities, let $IR(\mathcal{M}^m)=2|\mathcal{M}^m|=|\mathcal{M}_b^m|+|\mathcal{M}_s^m|$ be the number of inequalities from IR conditions.

Table \ref{tb:num_inequalities} summarizes the above results and illustrates a useful configuration of pairwise inequalities. Note that $_{|\mathcal{M}^m|+|\mathcal{N}_b^u|+|\mathcal{N}_s^u|}P_2$ and $_{|\mathcal{M}^m|}P_2$ include the same inequalities in $IR(\mathcal{M}^m)$. Section \ref{sec:identification_of_matching_cost} uses the result to demonstrate a necessary modification for identifying matching costs.

\begin{table}[!ht]
\caption{\textbf{The minimum number of total inequalities.}}
\centering
\begin{tabular}{|c|c|c|c| } 
 \hline
 \multicolumn{2}{|c|}{}&\multicolumn{2}{|c|}{Unmatched} \\ \cline{3-4}
 \multicolumn{2}{|c|}{}& Available & Unavailable \\ 
 \hline
 \multirow{2}{*}{ Transfer }&Available & $_{|\mathcal{M}^m|+|\mathcal{N}_b^u|+|\mathcal{N}_s^u|}P_2$ & $_{|\mathcal{M}^m|}P_2$ \\ \cline{2-4}
 &Unavailable & $_{|\mathcal{M}^m|+|\mathcal{N}_b^u|+|\mathcal{N}_s^u|}C_2+IR(\mathcal{M}^m)$ & $_{|\mathcal{M}^m|}C_2+IR(\mathcal{M}^m)$ \\ 
 \hline
\end{tabular}
\label{tb:num_inequalities}
\end{table}

\subsection{Matching maximum score estimator}\label{subsec:matching_maximum_score_estimator}

\cite{fox2010qe} proposes a maximum score
estimator using Inequality \eqref{eq:inequality_with_transfer} or \eqref{eq:inequality_without_transfer}. The maximum score estimator is consistent if the model satisfies a rank order property, i.e., the probability of observing matched pairs is larger than the probability of observing swapped matched pairs. I specify $f(b,s)$ as a parametric form $f(b,s|X,\beta)$ where $X$ is a vector of observed characteristics of all buyers and sellers and $\beta$ is a vector of parameters. Given $X$, one can estimate $\beta$ without IR conditions by maximizing the following objective function:
\begin{align}
    Q(\beta)=\begin{cases}
         \sum_{(b,s)\in \mathcal{M}} \sum_{(b',s')\in \mathcal{M},(b',s')\neq (b,s)} \mathbbm{1}[f(b,s|X,\beta)-f(b,s'|X,\beta)\ge p_{b,s}-p_{b',s'}]\\
         \quad\quad\quad\quad\quad\quad\quad\quad\quad\quad \text{if transfer data is available},\\
         \sum_{(b,s)\in \mathcal{M}} \sum_{(b',s')\in \mathcal{M},(b',s')\neq (b,s)} \mathbbm{1}[f(b,s|X,\beta)+ f(b',s'|X,\beta)\ge f(b,s'|X,\beta)+f(b',s|X,\beta)]\\
         \quad\quad\quad\quad\quad\quad\quad\quad\quad\quad \text{otherwise},
    \end{cases}\label{eq:score_function}
\end{align}
where $\mathbbm{1}[\cdot]$ is an indicator function. If unmatched data is unavailable, $\mathcal{M}$ is replaced with $\mathcal{M}^m$. If IR conditions are included, $Q(\beta)$ is modified to the following:
\begin{align}
    \tilde{Q}(\beta)=Q(\beta) + \lambda\cdot\sum_{(b,s)\in \mathcal{M}^m}  \mathbbm{1}[f(b,s|X,\beta)\ge 0], \quad \lambda \ge 1,\label{eq:score_function_with_IR}
\end{align}
where $\lambda$ is the importance weight of IR conditions. If $\lambda$ is larger, the importance of the IR condition term is larger for the evaluation of $\tilde{Q}(\beta)$. If the transfer data are available, the correction term is redundant as in Table \ref{tb:num_inequalities}, but it does not affect the search for the maximizer of $\tilde{Q}(\beta)$. Section \ref{sec:identification_of_matching_cost} investigates the importance of $\lambda$.

\section{IR conditions identify the coefficient parameter of a constant}\label{sec:identification_of_matching_cost}

The estimation of fixed costs is one of the fundamental tasks in structural empirical studies. Suppose that researchers want to estimate an additive separable matching cost in a single market. Let $c$ be the matching cost and assume $c<0$ for exposition.\footnote{If researchers expect $\beta X_bX_s$ to be negative, i.e., the market to generate negative assortative matchings and want to investigate a subsidy effect inducing matchings, they can assume $c>0$ as the subsidy effect. The logic is the same in the main text.} Following the literature, I specify $f(b,s)$ as follows:
\begin{align}
    f(b,s) = \beta X_b X_s + c \cdot\mathbbm{1}[b\neq \emptyset \text{ or } s\neq \emptyset],\label{eq:with_matching_cost}
\end{align}
where $X_b$ and $X_s$ are vectors of continuous observable characteristics for buyer $b$ and seller $s$ and $\beta$ is a vector of parameters. Note that hypothetical matching cost $c$ exists for unmatched pairs. Then, Result \ref{prop:identification_of_matching_cost} holds. 

\begin{proposition}\label{prop:identification_of_matching_cost}
Suppose that the matching joint production function is specified as \eqref{eq:with_matching_cost}. Then,
\begin{enumerate}
    \renewcommand{\labelenumi}{(\roman{enumi})}
    \item by using pairwise inequalities based only on matched pairs, $c$ cannot be identified via $Q(\cdot)$;
    \item in addition to (i), even if transfer data is available, $c$ cannot be identified via $Q(\cdot)$;
    \item in addition to (i), even if unmatched data is available, the lower bound of $c$ cannot be identified without IR conditions;
    \item in addition to (iii), $c$ can be identified only when IR conditions are used with a sufficiently large importance weight via $\tilde{Q}(\beta)$, whether the transfer data is included or not. 
\end{enumerate}
\end{proposition}

Note that $(iv)$ of Result \ref{prop:identification_of_matching_cost} does not reveal how large the importance weight $\lambda$ should be. The appropriate weight of $\lambda$ depends on the sample sizes of matched and unmatched agents as a tuning parameter. The working paper version provides detailed numerical experiments to show the necessity of the weight.




\section{Conclusion}\label{sec:conclusion}
This article investigates the identification of a common fixed cost of matching in a TU matching game. For identification, IR conditions with a sufficiently large importance weight are necessary unless you have data of both unmatched agents and transfers. 
\paragraph{Acknowledgments}
I thank my advisor Jeremy Fox for his valuable advice. This research did not receive any specific grant from funding agencies in the public, commercial, or not-for-profit sectors. 


\singlespacing

\appendix
\section{Proof of results}
\paragraph{Proof of Result 1}
\begin{proof}
I demonstrate $(i)$ because other parts are analogously proven. $\mathcal{M}$ restores unmatched buyer $\tilde{b}$ and seller $\tilde{s}$ as $(\tilde{b},\emptyset)$ and $(\emptyset,\tilde{s})$, so the size of the set of unmatched pairs is $(|\mathcal{N}_b^u|+|\mathcal{N}_s^u|)$.
The number of possible combinations of $(b,s)\in \mathcal{M}$ and $(b',s')\in \mathcal{M}, (b',s')\neq(b,s)$ is $_{|\mathcal{M}^m|+|\mathcal{N}_b^u|+|\mathcal{N}_s^u|}C_2$ and each combination gives two inequalities as in Inequality \eqref{eq:inequality_with_transfer}. This gives 
\begin{align}
    2\cdot _{|\mathcal{M}^m|+|\mathcal{N}_b^u|+|\mathcal{N}_s^u|}C_2=2\frac{(|\mathcal{M}^m|+|\mathcal{N}_b^u|+|\mathcal{N}_s^u|)(|\mathcal{M}^m|+|\mathcal{N}_b^u|+|\mathcal{N}_s^u|-1)}{2!}=_{|\mathcal{M}^m|+|\mathcal{N}_b^u|+|\mathcal{N}_s^u|}P_2.
\end{align}
\end{proof}
\paragraph{Proof of Result 2}
\begin{proof}
To prove $(i)$, substituting \eqref{eq:with_matching_cost} into \eqref{eq:inequality_with_transfer} gives
\begin{align*}
    &\beta X_b X_s + c \cdot\mathbbm{1}[b\neq \emptyset \text{ or } s\neq \emptyset] + \beta X_{b'} X_{s'} + c \cdot\mathbbm{1}[b'\neq \emptyset \text{ or } s'\neq \emptyset]\\
    &\quad\ge \beta X_b X_{s'} + c \cdot\mathbbm{1}[b\neq \emptyset \text{ or } s'\neq \emptyset] + \beta X_{b'} X_s + c \cdot\mathbbm{1}[b'\neq \emptyset \text{ or } s\neq \emptyset],
\end{align*}
where the indicator functions must be 1 for matched pair $(b,s)$ and unrealized matched pair $(b,s')$ so that matching cost $c$ is canceled out. Similarly, $(ii)$ is also proved in the same way.

To prove $(iii)$, it is sufficient to consider pairwise inequalities based on an unmatched pair as the pairwise inequalities of matched pairs are shown above. With transfer data, by substituting \eqref{eq:with_matching_cost} into \eqref{eq:inequality_with_transfer}, the pairwise inequalities based on unmatched pairs denoted by $(\tilde{b},\emptyset)$ and $(\emptyset,\tilde{s})$ can be reduced to a single inequality as follows:
\begin{align*}
    0&\ge \beta X_{\tilde{b}} X_{\tilde{s}} + c - p_{\tilde{b},\tilde{s}},
\end{align*}
where $p_{\tilde{b},\tilde{s}}=0$. This only provides the upper bound of $c$ as $-\beta X_b X_s\ge c$. Without transfer data, the same inequality is derived using \eqref{eq:inequality_without_transfer}.

To prove $(iv)$, when transfer data is available, IR condition \eqref{eq:IR_condition} for matched pair $(b,s)$ gives
\begin{align*}
    \beta X_b X_s + c - p_{b,s}\ge 0,
\end{align*}
whereas, when transfer data is not available, IR condition \eqref{eq:IR_condition} for matched pair $(b,s)$ gives
\begin{align*}
    \beta X_b X_s + c \ge 0.
\end{align*}
Thus, IR condition \eqref{eq:IR_condition} provides the lower bound of $c$ as $c\ge -\beta X_b X_s+p_{b,s}$ or $c\ge -\beta X_b X_s$. 

Finally, I demonstrate the necessity of $\lambda$. In the modified objective function in \eqref{eq:score_function_with_IR}, the additional term
\begin{align*}
    \lambda\cdot\sum_{(b,s)\in \mathcal{M}^m}  1[\beta X_b X_s + c\ge 0],
\end{align*}
takes up to $\lambda\cdot |\mathcal{M}^m|$. However, $Q(\beta)$ of \eqref{eq:score_function_with_IR} can take up to either of the numbers in Table \ref{tb:num_inequalities} which are much larger than $|\mathcal{M}^m|$. In finite samples, the part does not need to achieve its perfect score, so some fractions of pairwise inequalities are not satisfied even at the maximizer of the objective function. This implies that, if $\lambda$ and $|\mathcal{M}^m|$ are small, the evaluation of IR conditions is dominated by the number of unsatisfied pairwise inequalities, which can be larger than $|\mathcal{M}^m|$. Thus, $\lambda$ must be large enough, corresponding to $|\mathcal{M}^m|$.
\end{proof}

\bibliographystyle{aer.bst}
\bibliography{00identification}

\newpage
\section{Numerical experiments (Not For Publication)}\label{sec:simulation_bound}

In Section \ref{sec:simulation_bound}, I examine the improvement of estimation accuracy via each type of inequality through the objective function of the matching maximum score estimator in a finite sample. See author's github page for replication code.

\subsection{Computation of a matching equilibrium }\label{subsec:computation}
Let $F(b,s)=f(b,s)+\varepsilon_{b,s}$ be the match value function of match $(b,s)$ where $\varepsilon_{b,s}$ is drawn from zero median distribution. The equilibrium one-to-one matching allocation $\{m(b,s)\}_{b\in\mathcal{N}_b,s\in\mathcal{N}_s}$ is calculated by the following linear programming problem proposed by \cite{shapley1971assignment}:\footnote{See \cite{galichon2018optimal} for the computational properties and recent results.}

\begin{align*}
    \max_{\{m(b,s)\}_{b\in\mathcal{N}_b,s\in\mathcal{N}_s}} &F(b,s)\cdot m(b,s)\\
    \text{s.t. } 0&\le \sum_{b\in\mathcal{N}_b}m(b,s)\le 1\\
    0&\le \sum_{s\in\mathcal{N}_s}m(b,s)\le 1\\
    0&\le m(b,s) \quad \forall b \in \mathcal{N}_b,\forall s \in \mathcal{N}_s
\end{align*}
where $F(b,s)$ is specified as some parametric form of observed characteristics of buyer $b$ and seller $s$ by researchers. The dual of this linear programming problem also gives equilibrium prices. I assume that unobservable hypothetical equilibrium prices for matchings with unmatched buyers and sellers equal to zero.

\subsection{Numerical results}\label{subsec:results}

We are interested in the improvement of identification and estimation accuracy by adding unmatched firms, transfer data, and IR conditions. Also, I investigate possibly negative assortative matching which may generate unmatched agents via binding IR conditions, which is not investigated by the previous study. In Section \ref{subsec:results}, I compare the finite sample performances of the four models in Table \ref{tb:num_inequalities}. For the exposition, I list the models below.
\begin{itemize}
    \item Model 1: with unmatched data and transfer data, denoted by $(U, T)$
    \item Model 2: without unmatched data but with transfer data, denoted by $(T)$
    \item Model 3: with unmatched data but without transfer data, denoted by $(U)$
    \item Model 4: without unmatched data and transfer data, denoted by $(None)$
\end{itemize}
and I compare each model with and without IR conditions.

For simplicity, I assume $|\mathcal{N}_b|=|\mathcal{N}_s|$ in a single matching market. I specify the market size as one of $\{10, 20, 30, 50, 100\}$. The relative sizes of unmatched agents, i.e., $|\mathcal{N}^u_b| $ and $|\mathcal{N}^u_s|$ depend on values of parameters. The error term $\varepsilon_{b,s}$ is i.i.d. drawn from $N(0,1)$. I examine the following two cases:

\begin{itemize}
    \item Case 1: I specify the joint production function as $f(b,s)=\beta_0 X_{b0} X_{s0}+\beta_1\cdot X_{b1} X_{s1} + \beta_2\cdot X_{b2} X_{s2}$ where $\beta_0=1$ for normalization, $\beta_1=0.5$, and $\beta_2 \in \{-3 -2, -1, 0, 1\}$ as true parameters. I assume that for $i\in\{b,s\}$
\begin{align*}
    \left(\begin{array}{c}
X_{i0} \\
X_{i1} \\
X_{i2}
\end{array}\right) \sim_{iid}N\left(\left(\begin{array}{c}
3 \\
3 \\
3
\end{array}\right), \left(\begin{array}{ccc}
1 & 0.25 & 0.25 \\
0.25 & 1 & 0.25 \\
0.25 & 0.25 & 1
\end{array}\right)\right).
\end{align*}
    For example, positive $\beta_2$ induces positive assortative matchings, whereas negative $\beta_2$ induces negative assortative matchings which may generate unmatched agents via binding IR conditions.
    \item Case 2: The model is the same as Case 1 except that I specify the joint production function as $f(b,s)=\beta_0 X_{b0} X_{s0}+\beta_1\cdot X_{b1} X_{s1} + \beta_2\cdot \kappa \cdot \mathbbm{1}[b\neq \emptyset \text{ or } s\neq \emptyset]$ where $\kappa$ is a positive constant adjusting the level. For example, if $\beta_2$ is negative, the term captures matching fixed costs common in the market. I assume that $\kappa=8$ for generating unmatched agents in simulations.\footnote{The value of positive constant $\kappa$ only affects matching outcomes through the data generating process. For example, in Case 2, observed interaction term $\beta_0X_{b0} X_{s0}+\beta_1X_{b1} X_{s1}$ is $13.5 ( = 1\times 3\times3+0.5\times 3\times3 )$ on average. Then, to create unmatched firms, $13.5-\kappa\beta_1+\varepsilon_{bs}$ needs to be at least negative for some matching pair $(b,s)$. Thus, researchers expect composite estimand $\kappa\beta_1$ to be large than $13.5$ in practical situations since $\kappa$ is unknown.}
\end{itemize}
 For each scenario, I simulate 100 datasets and estimate parameters for each dataset under the same importance weight of IR conditions. I fix the weight $\lambda=100$.\footnote{In Appendix \ref{subsec:importance_level_lambda}, I investigate how the importance weight of IR conditions affects identification of parameters in detail. }

\subsubsection{Case 1: including negative assortativeness.}

\begin{figure}[htbp]
 \begin{minipage}{0.44\hsize}
  \begin{center}
   \includegraphics[width=60mm]{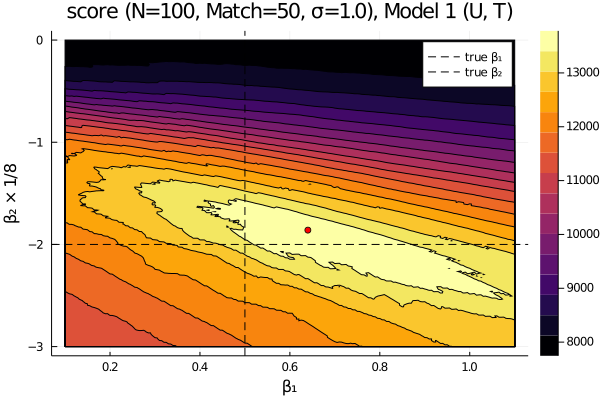}
  \end{center}
 \end{minipage}
 \begin{minipage}{0.44\hsize}
  \begin{center}
   \includegraphics[width=60mm]{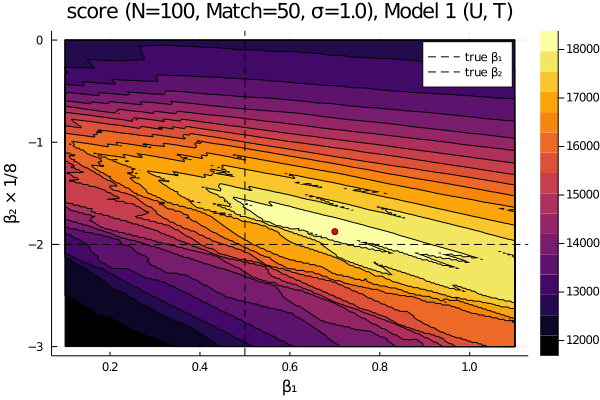}
  \end{center}
 \end{minipage}
 \\
 \begin{minipage}{0.44\hsize}
  \begin{center}
   \includegraphics[width=60mm]{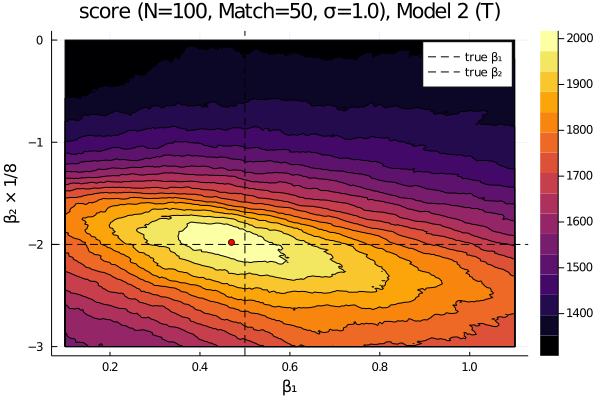}
  \end{center}
 \end{minipage}
 \begin{minipage}{0.44\hsize}
  \begin{center}
   \includegraphics[width=60mm]{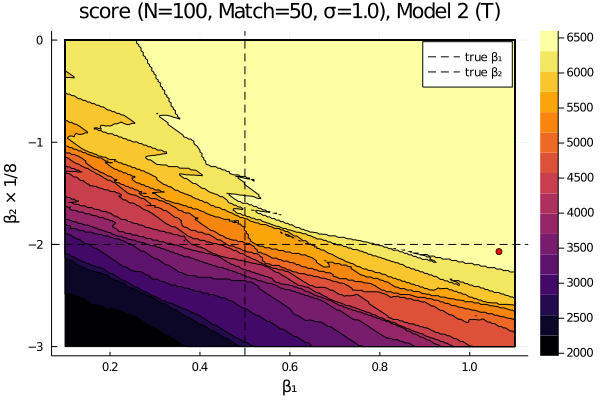}
  \end{center}
 \end{minipage}
 \\
 \begin{minipage}{0.44\hsize}
  \begin{center}
   \includegraphics[width=60mm]{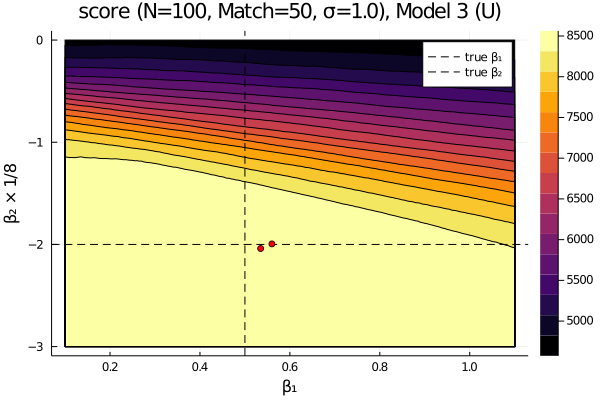}
  \end{center}
 \end{minipage}
 \begin{minipage}{0.44\hsize}
  \begin{center}
   \includegraphics[width=60mm]{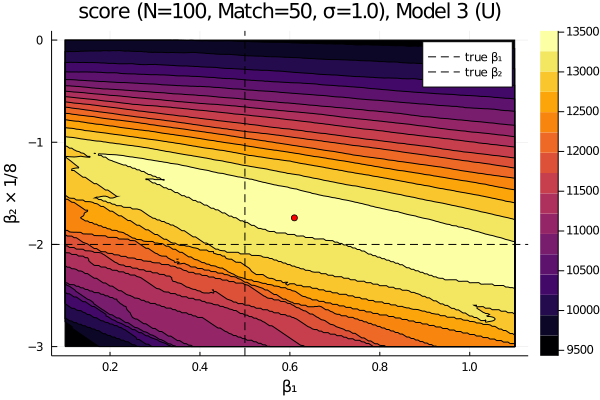}
  \end{center}
 \end{minipage}\\
 \begin{minipage}{0.44\hsize}
  \begin{center}
   \includegraphics[width=60mm]{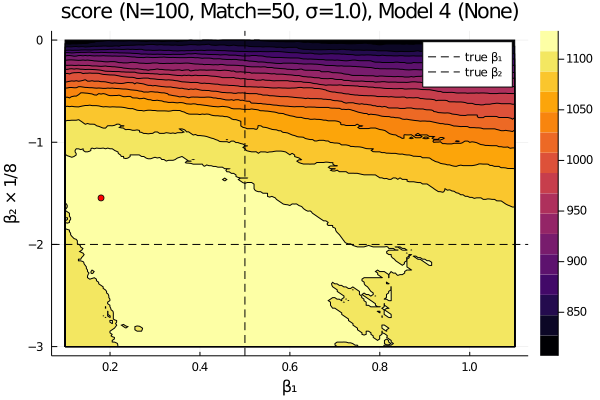}
  \end{center}
 \end{minipage}
 \begin{minipage}{0.44\hsize}
  \begin{center}
   \includegraphics[width=60mm]{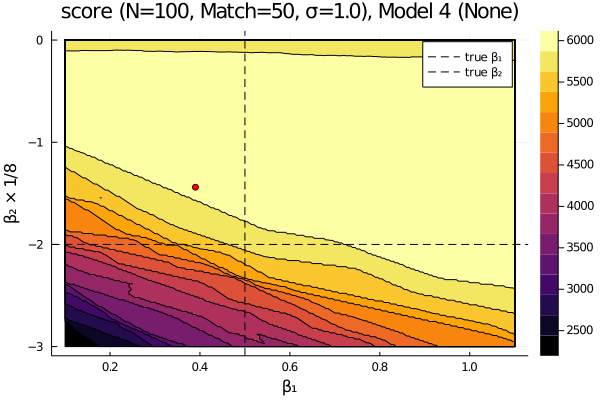}
  \end{center}
 \end{minipage}
 \caption{\textbf{Contour maps of the objective function of each model for Case 1 ($\lambda=100$).}}
 \begin{tablenotes}
\item[a]\textit{Note:} I use the same simulated single dataset with $\beta_0=+1,\beta_1=0.5,\beta_2=-2$ for all specifications. The only difference is the specification of the objective function. The left figures do not include the IR condition, whereas the right figures do include this. Models 1, 2, 3, and 4 are listed from top to bottom in order. The maxima of the objective function are plotted as red points.
\end{tablenotes}
 \label{fg:identified_set_case_1}
\end{figure}

Figure \ref{fg:identified_set_case_1} illustrates contour maps of the objective function over $\beta_1$ and $\beta_2$ under the same single simulated dataset across different models with and without IR conditions (left and right). Note that it captures only how the objective function is shaped and does not capture the identified set for set-identified cases so that parameters are point estimated. Models 1, 2, 3, and 4 are listed from top to bottom in order. I find that transfer data in itself is helpful for estimation regardless of including IR conditions and unmatched firms. It also shows that if researchers do not have transfer data, IR conditions trace out the lower bound of $\beta_2$ so that the objective function is sharpened. However, the estimated maximizer of the objective function can be slightly biased.

Tables \ref{tb:estimation_results_single_market_two_param_beta_without_dummy_without_IR} and \ref{tb:estimation_results_single_market_two_param_beta_without_dummy_with_IR} show the bias and Root Mean Squared Error (henceforth, RMSE) across four models with and without IR conditions. I find that for Case 1, including IR conditions decreases the bias and RMSE of $\beta_2$ in almost all specifications. In contrast, $\beta_1$ estimated with IR conditions is upward biased significantly for all models. Thus, for Case 1, there is no strong recommendation about whether IR conditions are included or not.

\subsubsection{Case 2: including matching fixed costs.}
\begin{figure}[htbp]
 \begin{minipage}{0.44\hsize}
  \begin{center}
   \includegraphics[width=60mm]{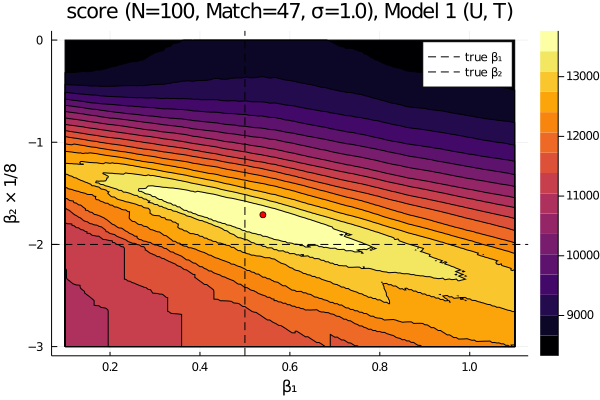}
  \end{center}
 \end{minipage}
 \begin{minipage}{0.44\hsize}
  \begin{center}
   \includegraphics[width=60mm]{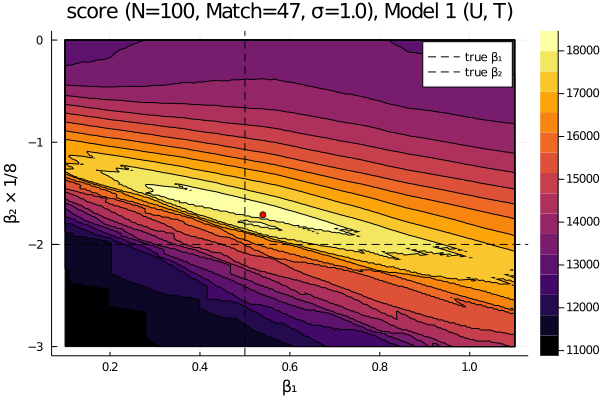}
  \end{center}
 \end{minipage}
 \\
 \begin{minipage}{0.44\hsize}
  \begin{center}
   \includegraphics[width=60mm]{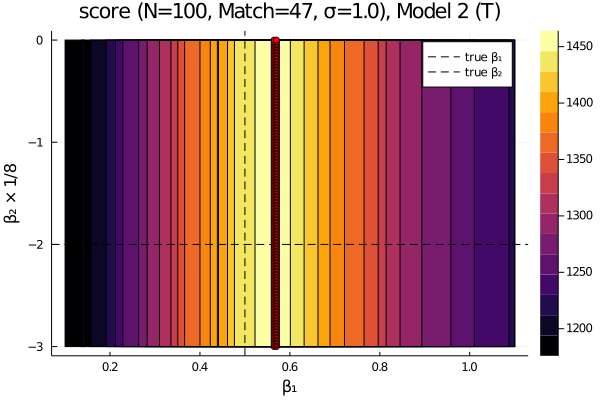}
  \end{center}
 \end{minipage}
 \begin{minipage}{0.44\hsize}
  \begin{center}
   \includegraphics[width=60mm]{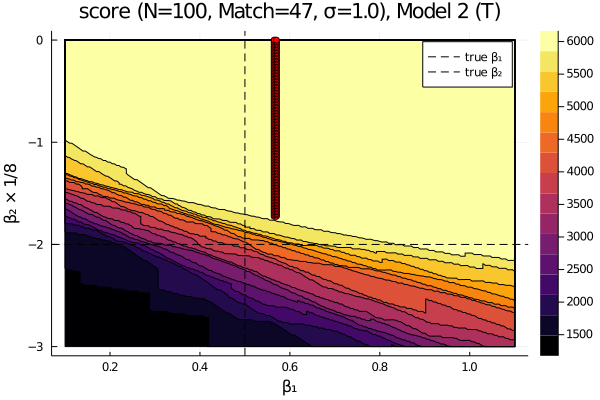}
  \end{center}
 \end{minipage}
 \\
 \begin{minipage}{0.44\hsize}
  \begin{center}
   \includegraphics[width=60mm]{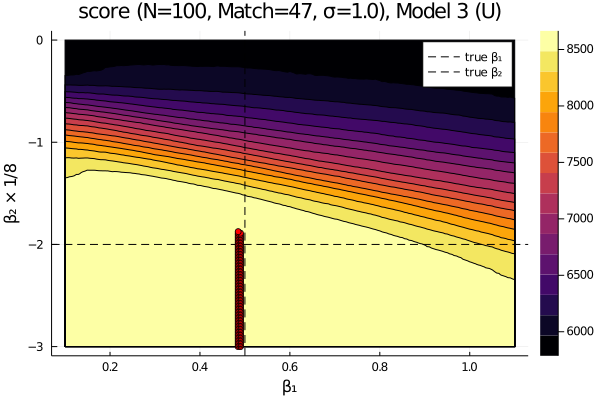}
  \end{center}
 \end{minipage}
 \begin{minipage}{0.44\hsize}
  \begin{center}
   \includegraphics[width=60mm]{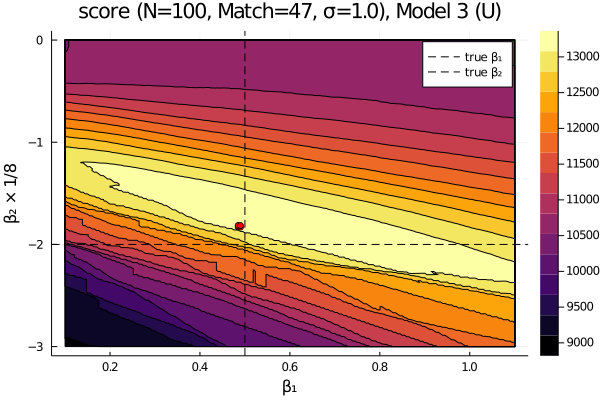}
  \end{center}
 \end{minipage}\\
 \begin{minipage}{0.44\hsize}
  \begin{center}
   \includegraphics[width=60mm]{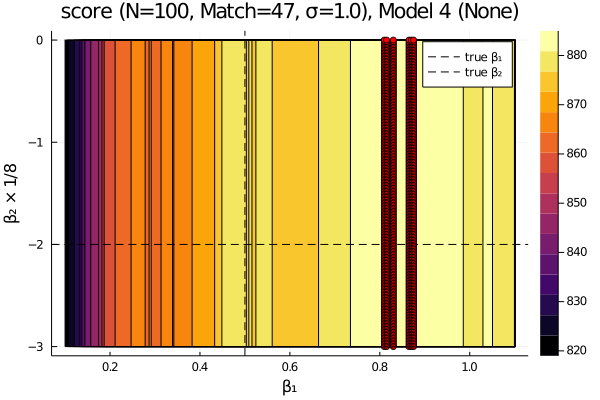}
  \end{center}
 \end{minipage}
 \begin{minipage}{0.44\hsize}
  \begin{center}
   \includegraphics[width=60mm]{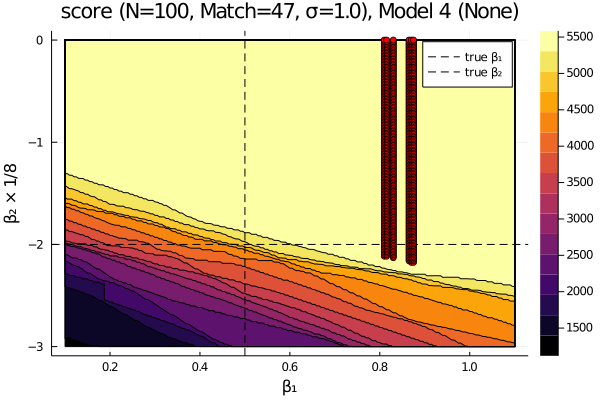}
  \end{center}
 \end{minipage}
 \caption{\textbf{Contour maps of the objective function of each model for Case 2 ($\lambda=100$).}}
 \begin{tablenotes}
\item[a]\textit{Note:} I use the same simulated single dataset with $\beta_0=+1,\beta_1=0.5,\beta_2=-2$ for all specifications. The only difference is the specification of the objective function. The left figures do not include the IR condition, whereas the right figures do include this. Models 1, 2, 3, and 4 are listed from top to bottom in order.  The maxima of the objective function are plotted as red points
\end{tablenotes}
 \label{fg:identified_set_case_2}
\end{figure}

Figure \ref{fg:identified_set_case_2} illustrates contour maps of the objective function over $\beta_1$ and $\beta_2$ under the same single simulated dataset across different models with and without IR conditions. The only difference from Figure \ref{fg:identified_set_case_1} is the specification of the post-merger matching joint production function $f(b,s)$ and generated data under the specification. First, the second and fourth left panels show that without IR conditions and unmatched data, $\beta_2$ cannot be identified completely. This is because pairwise inequalities based only on matched pairs cancel out the constant term regarding $\beta_2$. Also, the third left panel shows that even if unmatched data is available, without IR conditions, the lower bound of $\beta_2$ cannot be identified. This implies that pairwise inequalities based on unmatched firms and matched firms investigate only why unmatched firms do not match any other firms. The case is satisfied when the matching cost is infinitely large. In other words, pairwise inequalities based on unmatched firms and matched firms do not answer why matched firms can match their sellers. IR conditions answer the question.

The right panels in Figure \ref{fg:identified_set_case_2} show how IR conditions trace out the lower bound of $\beta_2$. In particular, the first and third right panels illustrate that if the data of unmatched firms is available, IR conditions help to identify the bound of $\beta_2$ regardless of the availability of transfer data. Since collecting data of unmatched agents is easier than collecting transfer data for each matching pair given the specified market definition, this finding is practically important.

Tables \ref{tb:estimation_results_single_market_two_param_beta_with_dummy_without_IR} and \ref{tb:estimation_results_single_market_two_param_beta_with_dummy_with_IR} confirm that a combination of IR conditions and unmatched data gives less biased estimates of $\beta_1$ and $\beta_2$ for almost all specifications. However, if less than 20 percent of the samples are unmatched as in the case of $\beta_2=-1$, the estimated $\beta_2$ is heavily biased upwardly because the information of unmatched choices is too less. Table \ref{tb:estimation_results_single_market_two_param_beta_with_dummy_with_IR_penalty_1} in Appendix \ref{subsec:importance_level_lambda} suggests that for such a case, $\lambda=1$ is an adequate weight. Table \ref{tb:estimation_results_single_market_two_param_beta_with_dummy_with_IR_safety_level_check_model_T} in Appendix \ref{subsec:how_many_unmatched} suggests that whether 30 percent of the full samples are unmatched or not is the practical threshold about whether researchers should use sufficiently high $\lambda$ or $\lambda=1$. The issue of the choice of $\lambda$ can be resolvable before implementing estimations given the data.

\begin{table}[!ht]
\caption{\textbf{Results of Monte Carlo simulations for Case 1 without IR.($\lambda=100$)}}
\begin{center}
{\tiny
  \begin{tabular}{@{\extracolsep{5pt}}lc|cccccc|lccccc}
\toprule 
 & Num of agents &  & 10 & 20 & 30 & 50 & 100 &  & 10 & 20 & 30 & 50 & 100 \\
$\beta_1$ &  &  &  &  &  &  &  & $\beta_2$ &  &  &  &  &  \\
\midrule 
0.5 & unmatched & Mean Num & 0.0 & 0.0 & 0.0 & 0.0 & 0.0 & 1.0 & 0.0 & 0.0 & 0.0 & 0.0 & 0.0 \\
 & U,T & Bias & 0.08 & 0.08 & 0.28 & 0.19 & 0.13 &  & 0.21 & 0.1 & 0.66 & 0.38 & 0.3 \\
 &  & RMSE & (0.5) & (0.35) & (0.68) & (0.39) & (0.27) &  & (0.58) & (0.39) & (1.39) & (0.72) & (0.63) \\
 & T & Bias & 0.17 & 0.08 & 0.28 & 0.21 & 0.13 &  & 0.25 & 0.09 & 0.66 & 0.42 & 0.32 \\
 &  & RMSE & (0.96) & (0.34) & (0.68) & (0.43) & (0.28) &  & (0.71) & (0.37) & (1.39) & (0.78) & (0.65) \\
 & U & Bias & 0.23 & 0.09 & 0.28 & 0.21 & 0.13 &  & 0.25 & 0.11 & 0.66 & 0.42 & 0.33 \\
 &  & RMSE & (1.11) & (0.36) & (0.68) & (0.43) & (0.28) &  & (0.7) & (0.4) & (1.39) & (0.78) & (0.65) \\
 & Non & Bias & 0.16 & 0.09 & 0.28 & 0.21 & 0.13 &  & 0.25 & 0.1 & 0.66 & 0.42 & 0.33 \\
 &  & RMSE & (0.96) & (0.35) & (0.68) & (0.43) & (0.28) &  & (0.71) & (0.39) & (1.39) & (0.78) & (0.65) \\
 &  &  &  &  &  &  &  &  &  &  &  &  &  \\
0.5 & unmatched & Mean Num & 0.01 & 0.0 & 0.0 & 0.0 & 0.03 & 0.0 & 0.01 & 0.0 & 0.0 & 0.0 & 0.03 \\
 & U,T & Bias & 1.19 & 0.46 & 0.46 & 0.33 & 0.38 &  & 0.21 & 0.04 & -0.0 & 0.02 & -0.01 \\
 &  & RMSE & (2.76) & (0.96) & (0.87) & (0.59) & (0.55) &  & (2.14) & (0.44) & (0.31) & (0.21) & (0.17) \\
 & T & Bias & 1.26 & 0.43 & 0.46 & 0.35 & 0.38 &  & 0.19 & 0.04 & -0.01 & 0.01 & -0.0 \\
 &  & RMSE & (2.82) & (0.91) & (0.87) & (0.6) & (0.55) &  & (2.17) & (0.45) & (0.31) & (0.21) & (0.17) \\
 & U & Bias & 1.23 & 0.43 & 0.46 & 0.36 & 0.39 &  & 0.19 & 0.03 & -0.01 & 0.01 & -0.01 \\
 &  & RMSE & (2.8) & (0.91) & (0.87) & (0.61) & (0.55) &  & (2.17) & (0.45) & (0.31) & (0.21) & (0.17) \\
 & Non & Bias & 1.26 & 0.43 & 0.46 & 0.36 & 0.39 &  & 0.2 & 0.03 & -0.01 & 0.01 & -0.01 \\
 &  & RMSE & (2.82) & (0.91) & (0.87) & (0.61) & (0.55) &  & (2.17) & (0.45) & (0.31) & (0.21) & (0.17) \\
 &  &  &  &  &  &  &  &  &  &  &  &  &  \\
0.5 & unmatched & Mean Num & 1.69 & 3.08 & 4.21 & 6.77 & 13.29 & -1.0 & 1.69 & 3.08 & 4.21 & 6.77 & 13.29 \\
 & U,T & Bias & 0.81 & 0.34 & 0.2 & 0.21 & 0.15 &  & -1.28 & -0.5 & -0.26 & -0.34 & -0.27 \\
 &  & RMSE & (1.75) & (0.65) & (0.43) & (0.38) & (0.31) &  & (2.04) & (0.79) & (0.49) & (0.58) & (0.44) \\
 & T & Bias & 0.57 & 0.38 & 0.17 & 0.22 & 0.14 &  & -1.17 & -0.57 & -0.26 & -0.3 & -0.25 \\
 &  & RMSE & (1.54) & (1.04) & (0.42) & (0.39) & (0.3) &  & (1.98) & (1.15) & (0.61) & (0.58) & (0.42) \\
 & U & Bias & 0.87 & 0.35 & 0.19 & 0.19 & 0.15 &  & -1.34 & -0.53 & -0.25 & -0.31 & -0.27 \\
 &  & RMSE & (1.86) & (0.67) & (0.44) & (0.37) & (0.3) &  & (2.18) & (0.84) & (0.5) & (0.56) & (0.45) \\
 & Non & Bias & 0.6 & 0.38 & 0.18 & 0.22 & 0.14 &  & -1.21 & -0.58 & -0.26 & -0.31 & -0.27 \\
 &  & RMSE & (1.56) & (1.04) & (0.44) & (0.39) & (0.3) &  & (2.03) & (1.15) & (0.6) & (0.59) & (0.44) \\
 &  &  &  &  &  &  &  &  &  &  &  &  &  \\
0.5 & unmatched & Mean Num & 5.68 & 10.65 & 15.73 & 25.47 & 50.12 & -2.0 & 5.68 & 10.65 & 15.73 & 25.47 & 50.12 \\
 & U,T & Bias & 0.31 & 0.15 & 0.12 & 0.14 & 0.11 &  & -1.43 & -0.63 & -0.5 & -0.37 & -0.27 \\
 &  & RMSE & (1.42) & (0.41) & (0.33) & (0.3) & (0.3) &  & (2.48) & (0.99) & (0.8) & (0.64) & (0.53) \\
 & T & Bias & 0.49 & 0.44 & 0.31 & 0.27 & 0.09 &  & -1.52 & -0.92 & -0.71 & -0.67 & -0.21 \\
 &  & RMSE & (2.41) & (1.48) & (1.03) & (0.67) & (0.24) &  & (2.74) & (1.92) & (1.58) & (1.56) & (0.6) \\
 & U & Bias & 0.28 & 0.12 & 0.09 & 0.12 & 0.08 &  & -1.38 & -0.58 & -0.48 & -0.36 & -0.24 \\
 &  & RMSE & (1.41) & (0.38) & (0.28) & (0.27) & (0.22) &  & (2.38) & (0.98) & (0.75) & (0.64) & (0.47) \\
 & Non & Bias & 0.35 & 0.46 & 0.31 & 0.27 & 0.09 &  & -1.27 & -0.91 & -0.71 & -0.69 & -0.22 \\
 &  & RMSE & (2.34) & (1.49) & (1.03) & (0.67) & (0.24) &  & (2.83) & (1.92) & (1.58) & (1.56) & (0.63) \\
 &  &  &  &  &  &  &  &  &  &  &  &  &  \\
0.5 & unmatched & Mean Num & 7.67 & 14.72 & 21.95 & 36.08 & 71.83 & -3.0 & 7.67 & 14.72 & 21.95 & 36.08 & 71.83 \\
 & U,T & Bias & -0.08 & 0.19 & 0.11 & 0.08 & 0.09 &  & -1.8 & -0.88 & -0.62 & -0.43 & -0.36 \\
 &  & RMSE & (2.04) & (0.51) & (0.35) & (0.29) & (0.2) &  & (2.76) & (1.64) & (1.27) & (0.82) & (0.64) \\
 & T & Bias & -0.26 & 0.45 & 0.25 & 0.07 & 0.21 &  & -0.48 & -0.91 & -0.88 & -0.17 & -0.79 \\
 &  & RMSE & (3.42) & (2.15) & (1.04) & (0.38) & (0.47) &  & (4.18) & (2.42) & (2.35) & (0.57) & (1.68) \\
 & U & Bias & -0.09 & 0.17 & 0.09 & 0.06 & 0.08 &  & -1.82 & -0.93 & -0.59 & -0.42 & -0.39 \\
 &  & RMSE & (2.04) & (0.51) & (0.32) & (0.27) & (0.2) &  & (2.76) & (1.71) & (1.22) & (0.82) & (0.63) \\
 & Non & Bias & -0.05 & 0.53 & 0.3 & 0.08 & 0.21 &  & -0.55 & -0.9 & -0.93 & -0.17 & -0.79 \\
 &  & RMSE & (3.58) & (2.35) & (1.2) & (0.37) & (0.47) &  & (4.11) & (2.41) & (2.42) & (0.57) & (1.68) \\
 &  &  &  &  &  &  &  &  &  &  &  &  &  \\
\hline 
\bottomrule 
\end{tabular}

\begin{tablenotes}
\textit{Note:}The objective function was numerically maximized using differential evolution (DE) algorithm in \texttt{BlackBoxOptim.jl} package. For the DE algorithm, I require setting the domain of parameters and the number of population seeds so that I fix the former to $[-10, 10]$ for parameters and the latter to 400. The reported point estimates are the best-found maxima. I omit the parameter $\beta_0$ that can take on only a finite number of values (here +1) and converge at an arbitrarily fast rate, they are superconsistent. 
\end{tablenotes}
}\end{center}
\label{tb:estimation_results_single_market_two_param_beta_without_dummy_without_IR}
\end{table}

\begin{table}[!ht]
\caption{\textbf{Results of Monte Carlo simulations for Case 1 with IR.($\lambda=100$)}}
\begin{center}
{\tiny
  \begin{tabular}{@{\extracolsep{5pt}}lc|cccccc|lccccc}
\toprule 
 & Num of agents &  & 10 & 20 & 30 & 50 & 100 &  & 10 & 20 & 30 & 50 & 100 \\
$\beta_1$ &  &  &  &  &  &  &  & $\beta_2$ &  &  &  &  &  \\
\midrule 
0.5 & unmatched & Mean Num & 0.0 & 0.0 & 0.0 & 0.0 & 0.0 & 1.0 & 0.0 & 0.0 & 0.0 & 0.0 & 0.0 \\
 & U,T & Bias & 0.18 & 0.07 & 0.3 & 0.21 & 0.12 &  & 0.35 & 0.16 & 0.72 & 0.43 & 0.31 \\
 &  & RMSE & (0.63) & (0.26) & (0.66) & (0.44) & (0.26) &  & (1.05) & (0.53) & (1.32) & (0.82) & (0.58) \\
 & T & Bias & 0.18 & 0.08 & 0.3 & 0.23 & 0.13 &  & 0.35 & 0.14 & 0.71 & 0.48 & 0.34 \\
 &  & RMSE & (0.63) & (0.27) & (0.65) & (0.47) & (0.28) &  & (1.05) & (0.49) & (1.32) & (0.88) & (0.61) \\
 & U & Bias & 0.18 & 0.08 & 0.3 & 0.23 & 0.12 &  & 0.34 & 0.17 & 0.72 & 0.48 & 0.33 \\
 &  & RMSE & (0.63) & (0.29) & (0.65) & (0.47) & (0.28) &  & (1.05) & (0.58) & (1.32) & (0.88) & (0.62) \\
 & Non & Bias & 0.18 & 0.08 & 0.3 & 0.23 & 0.12 &  & 0.34 & 0.17 & 0.72 & 0.48 & 0.33 \\
 &  & RMSE & (0.63) & (0.29) & (0.65) & (0.47) & (0.28) &  & (1.05) & (0.58) & (1.32) & (0.88) & (0.62) \\
 &  &  &  &  &  &  &  &  &  &  &  &  &  \\
0.5 & unmatched & Mean Num & 0.01 & 0.0 & 0.0 & 0.0 & 0.02 & 0.0 & 0.01 & 0.0 & 0.0 & 0.0 & 0.02 \\
 & U,T & Bias & 0.62 & 0.57 & 0.64 & 0.45 & 0.41 &  & 0.27 & 0.13 & 0.0 & 0.01 & 0.03 \\
 &  & RMSE & (1.68) & (1.17) & (1.12) & (0.66) & (0.58) &  & (1.16) & (0.56) & (0.27) & (0.21) & (0.19) \\
 & T & Bias & 0.59 & 0.59 & 0.66 & 0.45 & 0.42 &  & 0.27 & 0.15 & 0.0 & 0.0 & 0.04 \\
 &  & RMSE & (1.67) & (1.18) & (1.13) & (0.67) & (0.59) &  & (1.16) & (0.56) & (0.28) & (0.22) & (0.19) \\
 & U & Bias & 0.61 & 0.59 & 0.66 & 0.45 & 0.41 &  & 0.27 & 0.15 & 0.0 & -0.01 & 0.03 \\
 &  & RMSE & (1.69) & (1.18) & (1.13) & (0.67) & (0.59) &  & (1.16) & (0.56) & (0.28) & (0.22) & (0.19) \\
 & Non & Bias & 0.61 & 0.6 & 0.66 & 0.45 & 0.41 &  & 0.27 & 0.14 & 0.0 & -0.01 & 0.03 \\
 &  & RMSE & (1.69) & (1.18) & (1.13) & (0.67) & (0.59) &  & (1.16) & (0.56) & (0.28) & (0.22) & (0.19) \\
 &  &  &  &  &  &  &  &  &  &  &  &  &  \\
0.5 & unmatched & Mean Num & 1.72 & 3.17 & 4.33 & 6.67 & 12.68 & -1.0 & 1.72 & 3.17 & 4.33 & 6.67 & 12.68 \\
 & U,T & Bias & 1.2 & 0.8 & 0.59 & 0.49 & 0.45 &  & -0.54 & -0.2 & -0.08 & 0.06 & 0.06 \\
 &  & RMSE & (2.13) & (1.25) & (0.85) & (0.72) & (0.61) &  & (1.15) & (0.58) & (0.41) & (0.3) & (0.24) \\
 & T & Bias & 1.31 & 0.71 & 0.62 & 0.5 & 0.49 &  & -0.57 & -0.14 & -0.03 & 0.1 & 0.08 \\
 &  & RMSE & (2.42) & (1.16) & (0.91) & (0.73) & (0.67) &  & (1.31) & (0.54) & (0.43) & (0.34) & (0.26) \\
 & U & Bias & 1.21 & 0.77 & 0.57 & 0.5 & 0.48 &  & -0.55 & -0.19 & -0.08 & 0.05 & 0.05 \\
 &  & RMSE & (2.13) & (1.23) & (0.85) & (0.75) & (0.66) &  & (1.16) & (0.57) & (0.4) & (0.32) & (0.24) \\
 & Non & Bias & 1.22 & 0.71 & 0.62 & 0.5 & 0.48 &  & -0.52 & -0.15 & -0.03 & 0.09 & 0.08 \\
 &  & RMSE & (2.21) & (1.16) & (0.91) & (0.73) & (0.66) &  & (1.25) & (0.55) & (0.42) & (0.35) & (0.26) \\
 &  &  &  &  &  &  &  &  &  &  &  &  &  \\
0.5 & unmatched & Mean Num & 5.6 & 10.6 & 15.9 & 25.38 & 50.44 & -2.0 & 5.6 & 10.6 & 15.9 & 25.38 & 50.44 \\
 & U,T & Bias & 0.82 & 0.52 & 0.38 & 0.38 & 0.36 &  & -1.16 & -0.32 & -0.12 & -0.05 & 0.02 \\
 &  & RMSE & (2.0) & (0.89) & (0.7) & (0.61) & (0.58) &  & (2.28) & (0.9) & (0.59) & (0.53) & (0.57) \\
 & T & Bias & 1.21 & 0.98 & 0.51 & 0.53 & 0.4 &  & -0.72 & -0.37 & 0.08 & 0.07 & 0.33 \\
 &  & RMSE & (2.66) & (1.82) & (0.94) & (0.85) & (0.62) &  & (2.55) & (1.63) & (0.61) & (0.67) & (0.55) \\
 & U & Bias & 0.76 & 0.51 & 0.38 & 0.4 & 0.33 &  & -1.07 & -0.3 & -0.13 & -0.05 & 0.07 \\
 &  & RMSE & (1.95) & (0.87) & (0.65) & (0.62) & (0.51) &  & (2.21) & (0.87) & (0.57) & (0.54) & (0.44) \\
 & Non & Bias & 1.09 & 0.98 & 0.51 & 0.52 & 0.38 &  & -0.56 & -0.37 & 0.1 & 0.07 & 0.33 \\
 &  & RMSE & (2.58) & (1.82) & (0.94) & (0.85) & (0.6) &  & (2.62) & (1.63) & (0.63) & (0.67) & (0.55) \\
 &  &  &  &  &  &  &  &  &  &  &  &  &  \\
0.5 & unmatched & Mean Num & 7.62 & 15.01 & 22.0 & 36.08 & 71.75 & -3.0 & 7.62 & 15.01 & 22.0 & 36.08 & 71.75 \\
 & U,T & Bias & 0.45 & 0.35 & 0.33 & 0.28 & 0.26 &  & -1.71 & -0.48 & -0.28 & -0.01 & 0.05 \\
 &  & RMSE & (2.24) & (0.6) & (0.63) & (0.53) & (0.53) &  & (2.83) & (1.25) & (1.11) & (0.72) & (0.83) \\
 & T & Bias & 1.22 & 1.17 & 0.85 & 0.64 & 0.36 &  & 0.23 & 0.24 & -0.33 & 0.15 & 0.36 \\
 &  & RMSE & (3.15) & (2.29) & (1.47) & (1.09) & (0.61) &  & (4.12) & (2.93) & (1.76) & (1.42) & (0.81) \\
 & U & Bias & 0.4 & 0.33 & 0.29 & 0.26 & 0.21 &  & -1.64 & -0.46 & -0.23 & 0.01 & 0.13 \\
 &  & RMSE & (2.23) & (0.61) & (0.51) & (0.52) & (0.38) &  & (2.77) & (1.27) & (0.91) & (0.7) & (0.6) \\
 & Non & Bias & 1.41 & 1.21 & 0.87 & 0.65 & 0.34 &  & 0.26 & 0.23 & -0.35 & 0.14 & 0.38 \\
 &  & RMSE & (3.33) & (2.4) & (1.49) & (1.12) & (0.61) &  & (4.09) & (2.95) & (1.77) & (1.42) & (0.81) \\
 &  &  &  &  &  &  &  &  &  &  &  &  &  \\
\hline 
\bottomrule 
\end{tabular}

\begin{tablenotes}
\textit{Note:}The objective function was numerically maximized using differential evolution (DE) algorithm in \texttt{BlackBoxOptim.jl} package. For the DE algorithm, I require setting the domain of parameters and the number of population seeds so that I fix the former to $[-10, 10]$ for parameters and the latter to 400. The reported point estimates are the best-found maxima. I omit the parameter $\beta_0$ that can take on only a finite number of values (here +1) and converge at an arbitrarily fast rate, they are superconsistent. 
\end{tablenotes}
}\end{center}
\label{tb:estimation_results_single_market_two_param_beta_without_dummy_with_IR}
\end{table}

\begin{table}[!ht]
\caption{\textbf{Results of Monte Carlo simulations for Case 2 without IR.($\lambda=100$)}}
\begin{center}
{\tiny
  \begin{tabular}{@{\extracolsep{5pt}}lc|cccccc|lccccc}
\toprule 
 & Num of agents &  & 10 & 20 & 30 & 50 & 100 &  & 10 & 20 & 30 & 50 & 100 \\
$\beta_1$ &  &  &  &  &  &  &  & $\beta_2$ &  &  &  &  &  \\
\midrule 
0.5 & unmatched & Mean Num & 0.0 & 0.0 & 0.0 & 0.0 & 0.0 & 1.0 & 0.0 & 0.0 & 0.0 & 0.0 & 0.0 \\
 & U,T & Bias & 0.31 & 0.28 & 0.14 & 0.04 & 0.01 &  & -0.53 & -1.12 & -0.69 & -1.79 & -1.59 \\
 &  & RMSE & (1.64) & (0.84) & (0.46) & (0.29) & (0.16) &  & (5.73) & (5.12) & (5.76) & (5.88) & (5.88) \\
 & T & Bias & 0.41 & 0.31 & 0.14 & 0.04 & 0.0 &  & -0.49 & -1.17 & -0.75 & -1.71 & -1.82 \\
 &  & RMSE & (1.77) & (0.89) & (0.47) & (0.29) & (0.16) &  & (5.6) & (5.16) & (5.77) & (5.85) & (6.0) \\
 & U & Bias & 0.41 & 0.32 & 0.14 & 0.04 & 0.0 &  & -0.51 & -1.29 & -0.74 & -1.7 & -1.58 \\
 &  & RMSE & (1.77) & (0.89) & (0.47) & (0.29) & (0.16) &  & (5.57) & (5.18) & (5.75) & (5.86) & (5.98) \\
 & Non & Bias & 0.41 & 0.32 & 0.14 & 0.04 & 0.0 &  & -0.51 & -1.26 & -0.74 & -1.7 & -1.58 \\
 &  & RMSE & (1.77) & (0.89) & (0.47) & (0.29) & (0.16) &  & (5.57) & (5.14) & (5.75) & (5.86) & (5.98) \\
 &  &  &  &  &  &  &  &  &  &  &  &  &  \\
0.5 & unmatched & Mean Num & 0.01 & 0.0 & 0.0 & 0.0 & 0.02 & 0.0 & 0.01 & 0.0 & 0.0 & 0.0 & 0.02 \\
 & U,T & Bias & 0.51 & 0.24 & 0.14 & 0.04 & 0.01 &  & 0.97 & 0.2 & 0.3 & -0.86 & -0.94 \\
 &  & RMSE & (1.99) & (0.74) & (0.46) & (0.29) & (0.16) &  & (5.55) & (5.0) & (5.71) & (5.66) & (5.8) \\
 & T & Bias & 0.57 & 0.24 & 0.14 & 0.04 & 0.0 &  & 0.92 & 0.2 & 0.25 & -0.83 & -0.78 \\
 &  & RMSE & (2.06) & (0.74) & (0.47) & (0.29) & (0.16) &  & (5.5) & (5.0) & (5.73) & (5.68) & (5.81) \\
 & U & Bias & 0.6 & 0.24 & 0.14 & 0.04 & 0.0 &  & 1.01 & 0.12 & 0.26 & -0.71 & -0.56 \\
 &  & RMSE & (2.1) & (0.74) & (0.47) & (0.29) & (0.16) &  & (5.42) & (4.96) & (5.71) & (5.65) & (5.81) \\
 & Non & Bias & 0.51 & 0.24 & 0.14 & 0.04 & 0.0 &  & 0.94 & 0.12 & 0.26 & -0.71 & -0.56 \\
 &  & RMSE & (1.99) & (0.74) & (0.47) & (0.29) & (0.16) &  & (5.52) & (4.96) & (5.71) & (5.65) & (5.81) \\
 &  &  &  &  &  &  &  &  &  &  &  &  &  \\
0.5 & unmatched & Mean Num & 1.56 & 2.87 & 3.81 & 5.61 & 10.47 & -1.0 & 1.56 & 2.87 & 3.81 & 5.61 & 10.47 \\
 & U,T & Bias & 0.28 & 0.33 & 0.12 & 0.03 & 0.02 &  & -3.72 & -4.97 & -4.81 & -4.24 & -4.71 \\
 &  & RMSE & (1.45) & (1.07) & (0.43) & (0.21) & (0.15) &  & (5.35) & (5.64) & (5.57) & (5.05) & (5.53) \\
 & T & Bias & 0.28 & 0.38 & 0.14 & 0.04 & 0.03 &  & -3.08 & -3.14 & -2.92 & -2.55 & -2.81 \\
 &  & RMSE & (1.43) & (1.25) & (0.41) & (0.25) & (0.16) &  & (5.45) & (5.96) & (5.42) & (5.32) & (5.26) \\
 & U & Bias & 0.29 & 0.3 & 0.12 & 0.04 & 0.01 &  & -3.78 & -5.04 & -5.08 & -4.52 & -4.98 \\
 &  & RMSE & (1.45) & (1.06) & (0.44) & (0.22) & (0.15) &  & (5.36) & (5.66) & (5.77) & (5.23) & (5.68) \\
 & Non & Bias & 0.28 & 0.38 & 0.15 & 0.05 & 0.02 &  & -3.08 & -3.01 & -2.84 & -2.53 & -2.86 \\
 &  & RMSE & (1.43) & (1.25) & (0.48) & (0.25) & (0.15) &  & (5.45) & (5.98) & (5.56) & (5.31) & (5.32) \\
 &  &  &  &  &  &  &  &  &  &  &  &  &  \\
0.5 & unmatched & Mean Num & 5.57 & 11.0 & 16.11 & 26.31 & 51.08 & -2.0 & 5.57 & 11.0 & 16.11 & 26.31 & 51.08 \\
 & U,T & Bias & 0.26 & 0.07 & 0.05 & 0.04 & 0.02 &  & -4.29 & -4.56 & -4.45 & -4.37 & -4.34 \\
 &  & RMSE & (0.86) & (0.31) & (0.24) & (0.16) & (0.11) &  & (4.89) & (5.07) & (5.04) & (5.02) & (4.9) \\
 & T & Bias & 0.25 & 0.13 & 0.43 & 0.35 & 0.07 &  & -0.99 & 0.11 & -0.51 & 0.88 & 0.81 \\
 &  & RMSE & (2.0) & (1.47) & (1.66) & (0.94) & (0.28) &  & (5.73) & (5.32) & (5.64) & (5.84) & (5.36) \\
 & U & Bias & 0.28 & 0.05 & 0.03 & 0.03 & 0.02 &  & -4.46 & -4.69 & -4.63 & -4.61 & -4.52 \\
 &  & RMSE & (0.88) & (0.29) & (0.2) & (0.15) & (0.1) &  & (5.0) & (5.13) & (5.12) & (5.16) & (4.99) \\
 & Non & Bias & 0.25 & 0.14 & 0.43 & 0.35 & 0.07 &  & -0.99 & 0.06 & -0.61 & 0.72 & 0.8 \\
 &  & RMSE & (2.0) & (1.47) & (1.66) & (0.94) & (0.28) &  & (5.73) & (5.35) & (5.6) & (5.76) & (5.37) \\
 &  &  &  &  &  &  &  &  &  &  &  &  &  \\
0.5 & unmatched & Mean Num & 8.88 & 17.4 & 25.72 & 42.52 & 84.13 & -3.0 & 8.88 & 17.4 & 25.72 & 42.52 & 84.13 \\
 & U,T & Bias & -0.38 & 0.1 & 0.08 & 0.01 & 0.04 &  & -2.78 & -3.87 & -3.32 & -3.72 & -3.6 \\
 &  & RMSE & (2.38) & (0.55) & (0.4) & (0.19) & (0.18) &  & (4.1) & (4.37) & (3.91) & (4.28) & (4.08) \\
 & T & Bias & -0.04 & 0.3 & 0.23 & 0.08 & 0.91 &  & 0.41 & 2.7 & 1.65 & 3.34 & 2.82 \\
 &  & RMSE & (4.16) & (3.19) & (2.98) & (0.88) & (2.33) &  & (5.6) & (6.82) & (5.84) & (6.3) & (6.16) \\
 & U & Bias & -0.39 & 0.09 & 0.06 & 0.0 & 0.01 &  & -2.8 & -4.08 & -3.5 & -3.98 & -3.86 \\
 &  & RMSE & (2.37) & (0.54) & (0.36) & (0.18) & (0.11) &  & (4.12) & (4.52) & (4.01) & (4.42) & (4.27) \\
 & Non & Bias & -0.13 & 0.29 & 0.22 & 0.1 & 0.93 &  & 0.6 & 2.52 & 1.54 & 3.24 & 2.97 \\
 &  & RMSE & (4.24) & (3.19) & (2.98) & (0.88) & (2.33) &  & (5.6) & (6.74) & (5.71) & (6.3) & (6.17) \\
 &  &  &  &  &  &  &  &  &  &  &  &  &  \\
\hline 
\bottomrule 
\end{tabular}

\begin{tablenotes}
\textit{Note:}The objective function was numerically maximized using differential evolution (DE) algorithm in \texttt{BlackBoxOptim.jl} package. For the DE algorithm, I require setting the domain of parameters and the number of population seeds so that I fix the former to $[-10, 10]$ for parameters and the latter to 400. The reported point estimates are the best-found maxima. I omit the parameter $\beta_0$ that can take on only a finite number of values (here +1) and converge at an arbitrarily fast rate, they are superconsistent. 
\end{tablenotes}
}\end{center}
\label{tb:estimation_results_single_market_two_param_beta_with_dummy_without_IR}
\end{table}

\begin{table}[!ht]
\caption{\textbf{Results of Monte Carlo simulations for Case 2 with IR. ($\lambda=100$)}}
\begin{center}
{\tiny
  \begin{tabular}{@{\extracolsep{5pt}}lc|cccccc|lccccc}
\toprule 
 & Num of agents &  & 10 & 20 & 30 & 50 & 100 &  & 10 & 20 & 30 & 50 & 100 \\
$\beta_1$ &  &  &  &  &  &  &  & $\beta_2$ &  &  &  &  &  \\
\midrule 
0.5 & unmatched & Mean Num & 0.0 & 0.0 & 0.0 & 0.0 & 0.0 & 1.0 & 0.0 & 0.0 & 0.0 & 0.0 & 0.0 \\
 & U,T & Bias & 0.17 & 0.25 & 0.16 & 0.05 & 0.01 &  & 3.87 & 4.25 & 4.13 & 4.28 & 4.67 \\
 &  & RMSE & (0.96) & (0.78) & (0.49) & (0.28) & (0.16) &  & (4.84) & (5.06) & (5.07) & (5.16) & (5.43) \\
 & T & Bias & 0.17 & 0.26 & 0.16 & 0.05 & 0.01 &  & 3.87 & 4.32 & 4.19 & 4.29 & 4.66 \\
 &  & RMSE & (0.96) & (0.78) & (0.49) & (0.28) & (0.17) &  & (4.84) & (5.12) & (5.14) & (5.14) & (5.41) \\
 & U & Bias & 0.17 & 0.29 & 0.15 & 0.05 & 0.01 &  & 3.87 & 4.24 & 4.22 & 4.46 & 4.62 \\
 &  & RMSE & (0.96) & (0.86) & (0.49) & (0.28) & (0.17) &  & (4.84) & (5.05) & (5.15) & (5.29) & (5.38) \\
 & Non & Bias & 0.17 & 0.29 & 0.15 & 0.05 & 0.01 &  & 3.87 & 4.24 & 4.22 & 4.46 & 4.62 \\
 &  & RMSE & (0.96) & (0.86) & (0.49) & (0.28) & (0.17) &  & (4.84) & (5.05) & (5.15) & (5.29) & (5.38) \\
 &  &  &  &  &  &  &  &  &  &  &  &  &  \\
0.5 & unmatched & Mean Num & 0.01 & 0.0 & 0.0 & 0.0 & 0.02 & 0.0 & 0.01 & 0.0 & 0.0 & 0.0 & 0.02 \\
 & U,T & Bias & 0.63 & 0.26 & 0.16 & 0.05 & 0.01 &  & 4.79 & 5.33 & 5.13 & 5.41 & 5.56 \\
 &  & RMSE & (1.98) & (0.78) & (0.49) & (0.28) & (0.16) &  & (5.68) & (5.99) & (5.91) & (6.13) & (6.21) \\
 & T & Bias & 0.59 & 0.26 & 0.16 & 0.05 & 0.01 &  & 4.86 & 5.33 & 5.19 & 5.41 & 5.55 \\
 &  & RMSE & (1.95) & (0.78) & (0.49) & (0.28) & (0.16) &  & (5.72) & (5.99) & (5.98) & (6.14) & (6.21) \\
 & U & Bias & 0.63 & 0.29 & 0.15 & 0.05 & 0.01 &  & 4.84 & 5.24 & 5.22 & 5.46 & 5.55 \\
 &  & RMSE & (1.98) & (0.86) & (0.49) & (0.28) & (0.17) &  & (5.7) & (5.91) & (6.0) & (6.16) & (6.18) \\
 & Non & Bias & 0.63 & 0.29 & 0.15 & 0.05 & 0.01 &  & 4.84 & 5.24 & 5.22 & 5.46 & 5.55 \\
 &  & RMSE & (1.98) & (0.86) & (0.49) & (0.28) & (0.17) &  & (5.7) & (5.91) & (6.0) & (6.16) & (6.18) \\
 &  &  &  &  &  &  &  &  &  &  &  &  &  \\
0.5 & unmatched & Mean Num & 1.52 & 2.97 & 3.83 & 5.78 & 10.73 & -1.0 & 1.52 & 2.97 & 3.83 & 5.78 & 10.73 \\
 & U,T & Bias & 0.59 & 0.44 & 0.24 & 0.19 & 0.2 &  & 2.59 & 1.61 & 1.94 & 1.67 & 1.88 \\
 &  & RMSE & (1.59) & (1.16) & (0.59) & (0.44) & (0.38) &  & (4.73) & (3.42) & (3.76) & (3.54) & (3.72) \\
 & T & Bias & 0.63 & 0.41 & 0.23 & 0.14 & 0.15 &  & 2.94 & 2.83 & 2.9 & 3.52 & 3.15 \\
 &  & RMSE & (1.69) & (1.18) & (0.57) & (0.4) & (0.34) &  & (4.98) & (4.6) & (4.49) & (5.25) & (4.81) \\
 & U & Bias & 0.59 & 0.41 & 0.23 & 0.2 & 0.2 &  & 2.57 & 1.49 & 1.74 & 1.45 & 1.79 \\
 &  & RMSE & (1.59) & (1.18) & (0.58) & (0.44) & (0.39) &  & (4.7) & (3.22) & (3.46) & (3.19) & (3.61) \\
 & Non & Bias & 0.67 & 0.43 & 0.22 & 0.17 & 0.15 &  & 2.92 & 2.5 & 2.61 & 3.21 & 3.25 \\
 &  & RMSE & (1.73) & (1.24) & (0.57) & (0.41) & (0.34) &  & (5.05) & (4.32) & (4.23) & (4.94) & (4.98) \\
 &  &  &  &  &  &  &  &  &  &  &  &  &  \\
0.5 & unmatched & Mean Num & 5.5 & 11.15 & 16.41 & 26.41 & 51.16 & -2.0 & 5.5 & 11.15 & 16.41 & 26.41 & 51.16 \\
 & U,T & Bias & 0.55 & 0.33 & 0.24 & 0.29 & 0.35 &  & -0.44 & -0.08 & 0.1 & 0.05 & 0.01 \\
 &  & RMSE & (1.17) & (0.68) & (0.47) & (0.56) & (0.71) &  & (1.34) & (0.68) & (0.52) & (0.5) & (0.7) \\
 & T & Bias & 0.86 & 0.46 & 0.6 & 0.44 & 0.12 &  & 1.63 & 3.91 & 4.15 & 5.24 & 5.92 \\
 &  & RMSE & (2.18) & (1.71) & (1.83) & (1.02) & (0.37) &  & (4.3) & (5.93) & (5.63) & (6.6) & (7.07) \\
 & U & Bias & 0.55 & 0.29 & 0.21 & 0.29 & 0.28 &  & -0.45 & -0.04 & 0.13 & 0.07 & 0.1 \\
 &  & RMSE & (1.17) & (0.63) & (0.45) & (0.54) & (0.49) &  & (1.34) & (0.65) & (0.53) & (0.46) & (0.41) \\
 & Non & Bias & 0.82 & 0.46 & 0.65 & 0.43 & 0.13 &  & 1.74 & 3.94 & 4.14 & 5.4 & 5.86 \\
 &  & RMSE & (2.07) & (1.71) & (1.9) & (1.02) & (0.37) &  & (4.29) & (5.94) & (5.62) & (6.74) & (6.99) \\
 &  &  &  &  &  &  &  &  &  &  &  &  &  \\
0.5 & unmatched & Mean Num & 8.81 & 17.45 & 25.71 & 42.59 & 84.21 & -3.0 & 8.81 & 17.45 & 25.71 & 42.59 & 84.21 \\
 & U,T & Bias & -0.19 & 0.33 & 0.38 & 0.18 & 0.21 &  & -1.48 & -0.58 & -0.56 & -0.05 & -0.0 \\
 &  & RMSE & (2.38) & (0.82) & (0.74) & (0.43) & (0.44) &  & (2.72) & (1.63) & (1.32) & (0.68) & (0.68) \\
 & T & Bias & 1.26 & 1.28 & 0.42 & 0.13 & 1.4 &  & 1.42 & 4.63 & 5.38 & 6.08 & 5.82 \\
 &  & RMSE & (3.9) & (3.62) & (1.92) & (1.0) & (2.96) &  & (5.36) & (6.96) & (7.02) & (7.09) & (7.12) \\
 & U & Bias & -0.19 & 0.33 & 0.35 & 0.18 & 0.18 &  & -1.48 & -0.56 & -0.49 & -0.05 & 0.05 \\
 &  & RMSE & (2.38) & (0.87) & (0.72) & (0.43) & (0.37) &  & (2.72) & (1.69) & (1.24) & (0.69) & (0.57) \\
 & Non & Bias & 1.28 & 1.37 & 0.41 & 0.12 & 1.42 &  & 1.5 & 4.74 & 5.31 & 6.1 & 5.78 \\
 &  & RMSE & (3.87) & (3.67) & (1.92) & (1.0) & (2.97) &  & (5.4) & (7.02) & (6.93) & (7.1) & (7.08) \\
 &  &  &  &  &  &  &  &  &  &  &  &  &  \\
\hline 
\bottomrule 
\end{tabular}

\begin{tablenotes}
\textit{Note:}The objective function was numerically maximized using differential evolution (DE) algorithm in \texttt{BlackBoxOptim.jl} package. For the DE algorithm, I require setting the domain of parameters and the number of population seeds so that I fix the former to $[-10, 10]$ for parameters and the latter to 400. The reported point estimates are the best-found maxima. I omit the parameter $\beta_0$ that can take on only a finite number of values (here +1) and converge at an arbitrarily fast rate, they are superconsistent. 
\end{tablenotes}
}\end{center}
\label{tb:estimation_results_single_market_two_param_beta_with_dummy_with_IR}
\end{table}



\section{Other experiments (Not For Publication)}\label{sec:supplemental_results}

\subsection{How many unmatched agents should we include?}\label{subsec:how_many_unmatched}

I investigate the safe levels of the proportion of the number of unmatched agents to that of matched agents. In particular, I focus on Model 3, i.e., the case when unmatched data is available. Table \ref{tb:estimation_results_single_market_two_param_beta_with_dummy_with_IR_safety_level_check_model_T} shows that at least 30 \% of the full sample ($N=100$) should be unmatched agents. For example, if $\beta_2=-1.6$, then 33 \% of the full samples are not matched and the bias is 0.19. Table \ref{tb:estimation_results_single_market_two_param_beta_with_dummy_with_IR_safety_level_check_model_T} also suggests that if the number of unmatched agents is relatively small, identification of $\beta_2$ may fail and the estimate will be heavily biased upward. For example, if $\beta_2=-1.2$ and less than 20 \% of the full samples are unmatched, then the bias is 0.76. The case approaches to the case where unmatched data is not available, like Model 2 and 4.

\begin{table}[!ht]
\caption{\textbf{How many unmatched agents should we include?(Case 2 with IR, $\lambda=100$).}}
\begin{center}
{\tiny
  \begin{tabular}{@{\extracolsep{5pt}}lc|cccccc|lccccc}
\toprule 
 & Num of agents &  & 10 & 20 & 30 & 50 & 100 &  & 10 & 20 & 30 & 50 & 100 \\
$\beta_1$ &  &  &  &  &  &  &  & $\beta_2$ &  &  &  &  &  \\
\midrule 
0.5 & unmatched & Mean Num & 1.5 & 2.81 & 3.8 & 5.78 & 10.76 & -1.0 & 1.5 & 2.81 & 3.8 & 5.78 & 10.76 \\
 & U & Bias & 0.59 & 0.46 & 0.25 & 0.2 & 0.17 &  & 2.63 & 2.29 & 2.26 & 2.43 & 2.38 \\
 &  & RMSE & (1.63) & (1.22) & (0.56) & (0.47) & (0.32) &  & (4.49) & (4.15) & (4.03) & (4.23) & (3.93) \\
0.5 & unmatched & Mean Num & 1.93 & 3.38 & 4.87 & 7.38 & 13.71 & -1.1 & 1.93 & 3.38 & 4.87 & 7.38 & 13.71 \\
 & U & Bias & 0.6 & 0.44 & 0.33 & 0.22 & 0.19 &  & 2.25 & 1.93 & 1.57 & 2.66 & 1.98 \\
 &  & RMSE & (1.46) & (0.87) & (0.62) & (0.44) & (0.36) &  & (4.44) & (4.01) & (3.35) & (4.46) & (3.84) \\
0.5 & unmatched & Mean Num & 2.28 & 4.12 & 5.89 & 9.11 & 17.19 & -1.2 & 2.28 & 4.12 & 5.89 & 9.11 & 17.19 \\
 & U & Bias & 0.49 & 0.52 & 0.33 & 0.33 & 0.36 &  & 1.85 & 1.11 & 1.22 & 1.67 & 0.76 \\
 &  & RMSE & (1.59) & (1.2) & (0.55) & (0.52) & (0.52) &  & (4.08) & (3.0) & (2.93) & (3.46) & (2.24) \\
0.5 & unmatched & Mean Num & 2.52 & 4.96 & 7.02 & 11.03 & 20.59 & -1.3 & 2.52 & 4.96 & 7.02 & 11.03 & 20.59 \\
 & U & Bias & 0.6 & 0.49 & 0.33 & 0.35 & 0.29 &  & 1.5 & 1.06 & 0.77 & 1.29 & 0.53 \\
 &  & RMSE & (1.44) & (0.85) & (0.58) & (0.6) & (0.44) &  & (3.62) & (3.03) & (2.28) & (2.99) & (1.32) \\
0.5 & unmatched & Mean Num & 2.96 & 5.75 & 8.2 & 13.0 & 24.6 & -1.4 & 2.96 & 5.75 & 8.2 & 13.0 & 24.6 \\
 & U & Bias & 0.67 & 0.41 & 0.4 & 0.44 & 0.34 &  & 0.86 & 0.93 & 0.5 & 0.48 & 0.3 \\
 &  & RMSE & (1.49) & (0.82) & (0.72) & (0.64) & (0.52) &  & (2.92) & (2.68) & (1.61) & (1.58) & (0.42) \\
0.5 & unmatched & Mean Num & 3.32 & 6.57 & 9.51 & 15.18 & 28.66 & -1.5 & 3.32 & 6.57 & 9.51 & 15.18 & 28.66 \\
 & U & Bias & 0.84 & 0.49 & 0.43 & 0.35 & 0.31 &  & 0.05 & 0.59 & 0.51 & 0.33 & 0.28 \\
 &  & RMSE & (1.69) & (0.93) & (0.65) & (0.5) & (0.5) &  & (2.14) & (2.28) & (1.8) & (1.02) & (0.43) \\
0.5 & unmatched & Mean Num & 3.79 & 7.32 & 10.71 & 17.6 & 33.29 & -1.6 & 3.79 & 7.32 & 10.71 & 17.6 & 33.29 \\
 & U & Bias & 0.78 & 0.62 & 0.4 & 0.37 & 0.38 &  & -0.18 & 0.17 & 0.16 & 0.26 & 0.19 \\
 &  & RMSE & (1.39) & (0.96) & (0.69) & (0.61) & (0.54) &  & (1.82) & (1.58) & (0.7) & (0.99) & (0.37) \\
0.5 & unmatched & Mean Num & 4.24 & 8.23 & 12.1 & 19.67 & 37.79 & -1.7 & 4.24 & 8.23 & 12.1 & 19.67 & 37.79 \\
 & U & Bias & 0.66 & 0.53 & 0.46 & 0.39 & 0.39 &  & -0.39 & -0.03 & 0.05 & 0.21 & 0.14 \\
 &  & RMSE & (1.27) & (0.89) & (0.77) & (0.6) & (0.62) &  & (1.41) & (0.76) & (0.56) & (0.45) & (0.44) \\
0.5 & unmatched & Mean Num & 4.75 & 9.2 & 13.47 & 22.06 & 42.32 & -1.8 & 4.75 & 9.2 & 13.47 & 22.06 & 42.32 \\
 & U & Bias & 0.57 & 0.58 & 0.43 & 0.46 & 0.36 &  & -0.23 & -0.03 & 0.13 & 0.05 & 0.15 \\
 &  & RMSE & (0.97) & (0.91) & (0.77) & (0.71) & (0.54) &  & (1.54) & (1.39) & (0.91) & (0.52) & (0.42) \\
0.5 & unmatched & Mean Num & 5.12 & 10.2 & 14.87 & 24.3 & 46.62 & -1.9 & 5.12 & 10.2 & 14.87 & 24.3 & 46.62 \\
 & U & Bias & 0.7 & 0.57 & 0.47 & 0.43 & 0.39 &  & -0.45 & -0.14 & 0.05 & 0.03 & 0.07 \\
 &  & RMSE & (1.32) & (0.9) & (0.78) & (0.68) & (0.58) &  & (1.49) & (0.73) & (0.56) & (0.55) & (0.4) \\
0.5 & unmatched & Mean Num & 5.55 & 11.08 & 16.12 & 26.46 & 51.1 & -2.0 & 5.55 & 11.08 & 16.12 & 26.46 & 51.1 \\
 & U & Bias & 0.58 & 0.47 & 0.45 & 0.41 & 0.44 &  & -0.37 & -0.08 & 0.04 & 0.08 & -0.01 \\
 &  & RMSE & (1.04) & (0.75) & (0.75) & (0.6) & (0.65) &  & (1.22) & (0.64) & (0.59) & (0.45) & (0.45) \\
0.5 & unmatched & Mean Num & 6.05 & 11.86 & 17.49 & 28.55 & 55.25 & -2.1 & 6.05 & 11.86 & 17.49 & 28.55 & 55.25 \\
 & U & Bias & 0.72 & 0.55 & 0.51 & 0.33 & 0.31 &  & -0.66 & -0.27 & -0.11 & 0.1 & 0.1 \\
 &  & RMSE & (1.47) & (0.92) & (0.83) & (0.5) & (0.54) &  & (1.88) & (1.01) & (0.72) & (0.41) & (0.52) \\
0.5 & unmatched & Mean Num & 6.44 & 12.58 & 18.69 & 30.6 & 59.37 & -2.2 & 6.44 & 12.58 & 18.69 & 30.6 & 59.37 \\
 & U & Bias & 0.64 & 0.49 & 0.43 & 0.36 & 0.32 &  & -0.59 & -0.2 & -0.03 & 0.06 & 0.04 \\
 &  & RMSE & (1.35) & (0.84) & (0.71) & (0.56) & (0.5) &  & (1.81) & (0.91) & (0.65) & (0.56) & (0.48) \\
0.5 & unmatched & Mean Num & 6.91 & 13.35 & 19.78 & 32.49 & 63.34 & -2.3 & 6.91 & 13.35 & 19.78 & 32.49 & 63.34 \\
 & U & Bias & 0.71 & 0.38 & 0.47 & 0.38 & 0.32 &  & -0.79 & -0.09 & -0.13 & -0.01 & 0.04 \\
 &  & RMSE & (1.46) & (0.66) & (0.71) & (0.6) & (0.53) &  & (1.99) & (0.75) & (0.68) & (0.58) & (0.56) \\
0.5 & unmatched & Mean Num & 7.2 & 14.1 & 20.79 & 34.17 & 67.23 & -2.4 & 7.2 & 14.1 & 20.79 & 34.17 & 67.23 \\
 & U & Bias & 0.63 & 0.45 & 0.35 & 0.3 & 0.31 &  & -0.74 & -0.26 & -0.05 & 0.08 & 0.03 \\
 &  & RMSE & (1.4) & (0.84) & (0.61) & (0.57) & (0.53) &  & (2.04) & (1.14) & (0.71) & (0.63) & (0.61) \\
0.5 & unmatched & Mean Num & 7.46 & 14.79 & 21.7 & 35.69 & 70.88 & -2.5 & 7.46 & 14.79 & 21.7 & 35.69 & 70.88 \\
 & U & Bias & 0.49 & 0.47 & 0.33 & 0.33 & 0.24 &  & -0.6 & -0.34 & -0.03 & 0.03 & 0.09 \\
 &  & RMSE & (1.37) & (0.94) & (0.62) & (0.56) & (0.42) &  & (1.98) & (1.37) & (0.69) & (0.6) & (0.5) \\
0.5 & unmatched & Mean Num & 7.82 & 15.33 & 22.67 & 37.37 & 74.26 & -2.6 & 7.82 & 15.33 & 22.67 & 37.37 & 74.26 \\
 & U & Bias & 0.32 & 0.37 & 0.29 & 0.29 & 0.31 &  & -0.75 & -0.29 & -0.02 & 0.01 & -0.03 \\
 &  & RMSE & (1.31) & (0.79) & (0.59) & (0.51) & (0.52) &  & (1.98) & (1.27) & (0.71) & (0.59) & (0.62) \\
0.5 & unmatched & Mean Num & 8.11 & 15.99 & 23.57 & 38.96 & 77.06 & -2.7 & 8.11 & 15.99 & 23.57 & 38.96 & 77.06 \\
 & U & Bias & 0.34 & 0.45 & 0.27 & 0.3 & 0.19 &  & -0.8 & -0.37 & -0.02 & -0.03 & 0.17 \\
 &  & RMSE & (1.18) & (0.76) & (0.61) & (0.49) & (0.36) &  & (1.82) & (1.08) & (0.83) & (0.57) & (0.48) \\
0.5 & unmatched & Mean Num & 8.38 & 16.54 & 24.43 & 40.21 & 79.72 & -2.8 & 8.38 & 16.54 & 24.43 & 40.21 & 79.72 \\
 & U & Bias & 0.26 & 0.35 & 0.39 & 0.29 & 0.25 &  & -0.97 & -0.33 & -0.27 & 0.01 & 0.04 \\
 &  & RMSE & (1.32) & (0.67) & (0.68) & (0.49) & (0.44) &  & (2.1) & (1.09) & (0.94) & (0.58) & (0.59) \\
0.5 & unmatched & Mean Num & 8.63 & 17.01 & 25.25 & 41.41 & 82.31 & -2.9 & 8.63 & 17.01 & 25.25 & 41.41 & 82.31 \\
 & U & Bias & 0.26 & 0.43 & 0.37 & 0.31 & 0.32 &  & -1.43 & -0.58 & -0.3 & -0.06 & -0.08 \\
 &  & RMSE & (1.63) & (0.88) & (0.61) & (0.52) & (0.54) &  & (2.73) & (1.63) & (0.88) & (0.67) & (0.73) \\
0.5 & unmatched & Mean Num & 8.81 & 17.42 & 25.84 & 42.61 & 84.38 & -3.0 & 8.81 & 17.42 & 25.84 & 42.61 & 84.38 \\
 & U & Bias & -0.22 & 0.48 & 0.35 & 0.26 & 0.21 &  & -1.45 & -0.73 & -0.32 & -0.03 & 0.05 \\
 &  & RMSE & (2.38) & (1.01) & (0.64) & (0.49) & (0.39) &  & (2.72) & (1.81) & (1.0) & (0.69) & (0.54) \\
 &  &  &  &  &  &  &  &  &  &  &  &  &  \\
\hline 
\bottomrule 
\end{tabular}

\begin{tablenotes}
\textit{Note:}The objective function was numerically maximized using differential evolution (DE) algorithm in \texttt{BlackBoxOptim.jl} package. For DE algorithm, I require setting the domain of parameters and the number of population seeds so that I fix the former to $[-10, 10]$ for parameters and the latter to 400. The reported point estimates are the best-found maxima. I omit the parameter $\beta_0$ that can take on only a finite number of values (here +1) converge at an arbitrarily fast rate, they are superconsistent. 
\end{tablenotes}
}\end{center}
\label{tb:estimation_results_single_market_two_param_beta_with_dummy_with_IR_safety_level_check_model_T}
\end{table}

\subsection{Sensitivity of $\lambda$, the importance weight of IR conditions.}\label{subsec:importance_level_lambda}

\begin{figure}[htbp]
 \begin{minipage}{0.5\hsize}
  \begin{center}
   \includegraphics[width=60mm]{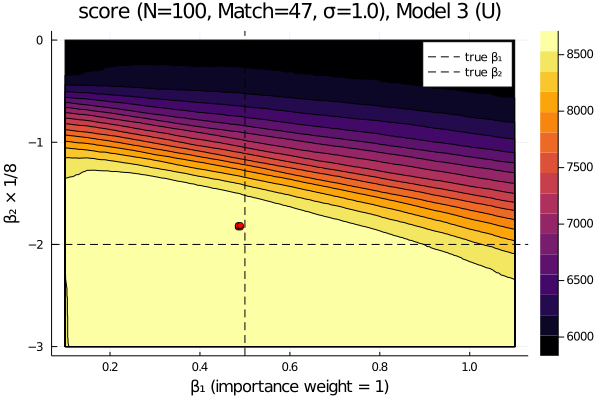}
  \end{center}
 \end{minipage}
 \begin{minipage}{0.5\hsize}
  \begin{center}
   \includegraphics[width=60mm]{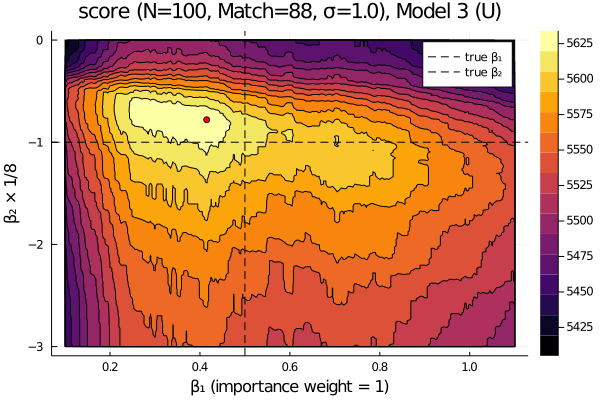}
  \end{center}
 \end{minipage}\\
 \begin{minipage}{0.5\hsize}
  \begin{center}
   \includegraphics[width=60mm]{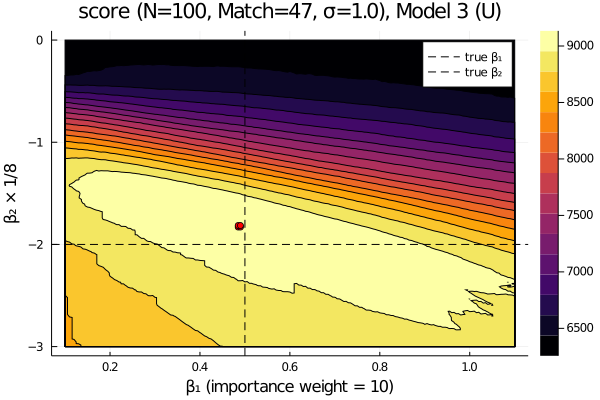}
  \end{center}
 \end{minipage}
 \begin{minipage}{0.5\hsize}
  \begin{center}
   \includegraphics[width=60mm]{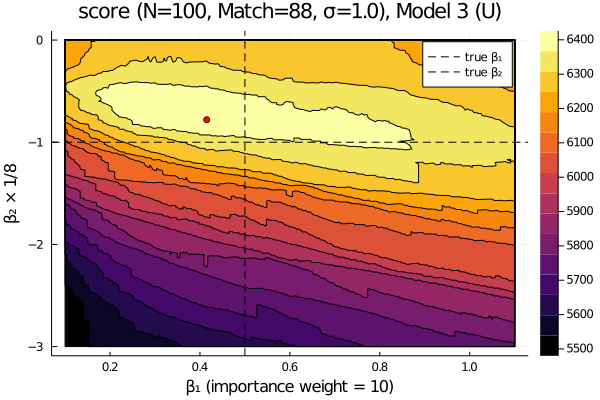}
  \end{center}
 \end{minipage}\\
 \begin{minipage}{0.5\hsize}
  \begin{center}
   \includegraphics[width=60mm]{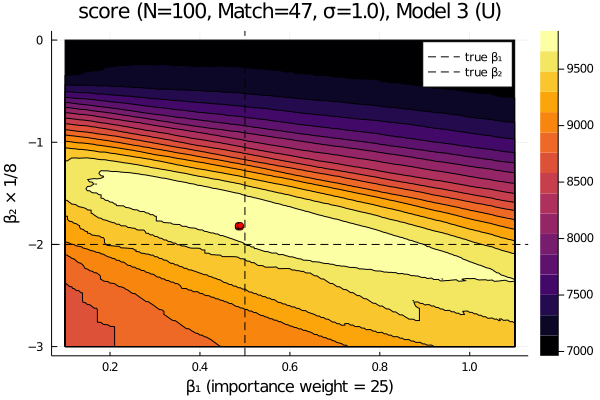}
  \end{center}
 \end{minipage}
 \begin{minipage}{0.5\hsize}
  \begin{center}
   \includegraphics[width=60mm]{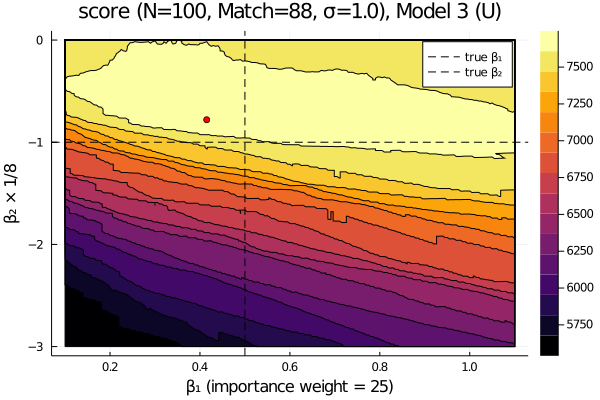}
  \end{center}
 \end{minipage}\\
 \begin{minipage}{0.5\hsize}
  \begin{center}
   \includegraphics[width=60mm]{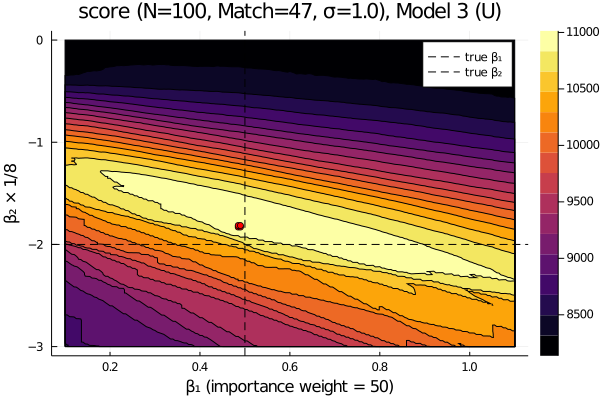}
  \end{center}
 \end{minipage}
 \begin{minipage}{0.5\hsize}
  \begin{center}
   \includegraphics[width=60mm]{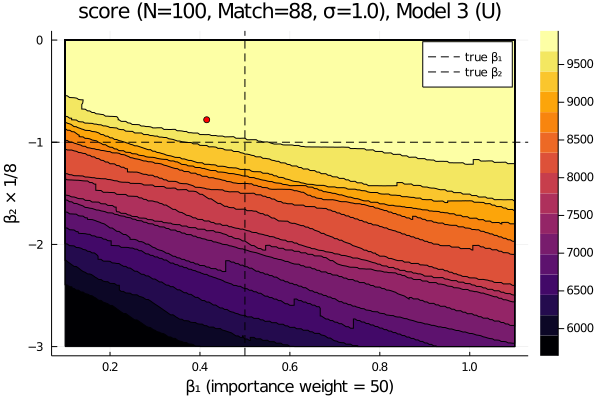}
  \end{center}
 \end{minipage}
 \caption{\textbf{Contour maps of the objective function of Model 3 for Case 2 under different $\lambda$ ($\beta_2=-2$ (left) and $\beta_2=-1$ (right)).}}
 \begin{tablenotes}
\item[a]\textit{Note:} I use Model 3 (i.e., $(U)$) under the same simulated single dataset with $\beta_0=+1,\beta_1=0.5$ and $\beta_2=-2$ (left) and $-1$ (right). The only difference is $\lambda$, the importance weight of IR conditions. I pick up $\lambda$ from $\{1,10,25,50\}$. The maximum point of the objective function is plotted. The maxima of the objective function are plotted as red points
\end{tablenotes}
 \label{fg:identified_set_case_2_penalty_level}
\end{figure}

Figure \ref{fg:identified_set_case_2_penalty_level} illustrates how the importance weight of IR conditions sharpens the objective function of $\beta_2$ for Model 3, that is, the case when unmatched data are available. The first left panel for $\beta_2=-2$ shows that even the lowest importance weight ($\lambda=1$) is helpful to obtain a maximizer point. Furthermore, the larger the importance weight is, the tighter the lower bound of $\beta_2$ is. This implies that DE algorithm is easier to find the maximizer under the larger importance weight. On the contrary, the first right panel of $\beta_2=-1$ shows that the lowest importance weight ($\lambda=1$) is enough for correction, which is consistent with Table \ref{tb:estimation_results_single_market_two_param_beta_with_dummy_with_IR_penalty_1}. Then, large $\lambda$ may loosen the upper bound of $\beta_2$. It should be noted that incorporating IR conditions with importance weight is not significantly harmful for estimations of all other parameters. Therefore, the researchers who want to estimate a matching cost must include IR conditions with the recommended importance weight in any case.

Tables \ref{tb:estimation_results_single_market_two_param_beta_with_dummy_with_IR_penalty_1}, \ref{tb:estimation_results_single_market_two_param_beta_with_dummy_with_IR_penalty_2}, \ref{tb:estimation_results_single_market_two_param_beta_with_dummy_with_IR_penalty_5}, \ref{tb:estimation_results_single_market_two_param_beta_with_dummy_with_IR_penalty_10}, \ref{tb:estimation_results_single_market_two_param_beta_with_dummy_with_IR_penalty_20} correspond to Table \ref{tb:estimation_results_single_market_two_param_beta_with_dummy_with_IR} with different importance weight $\lambda$. These suggest an adequate weight of $\lambda$. Concretely, if $\beta_2$ is too low and more than half of the samples are not matched, then small $\lambda$ induces the downward bias of $\beta_2$. This is because the correction term \eqref{eq:score_function_with_IR} is ignorable and the lower bound is loose when $|\mathcal{M}^m|$ is small. On the other hand, if $\beta_2$ is not low and more than 90 \% of sample are matched, then $\lambda=1$ gives unbiased estimates rather than large $\lambda$ because $|\mathcal{M}^m|$ is sufficiently large. As a numerical finding, whether 30 \% of the full samples are unmatched or not is the practical threshold about whether researchers should use sufficiently high $\lambda$ or $\lambda=1$.

\begin{table}[!ht]
\caption{\textbf{Case 2 with IR ($\lambda = 1$).}}
\begin{center}
{\tiny
  \begin{tabular}{@{\extracolsep{5pt}}lc|cccccc|lccccc}
\toprule 
 & Num of agents &  & 10 & 20 & 30 & 50 & 100 &  & 10 & 20 & 30 & 50 & 100 \\
$\beta_1$ &  &  &  &  &  &  &  & $\beta_2$ &  &  &  &  &  \\
\midrule 
0.5 & unmatched & Mean Num & 0.0 & 0.0 & 0.0 & 0.0 & 0.0 & 1.0 & 0.0 & 0.0 & 0.0 & 0.0 & 0.0 \\
 & U,T & Bias & 0.42 & 0.17 & 0.15 & 0.05 & 0.01 &  & 3.71 & 4.09 & 4.3 & 3.48 & 4.15 \\
 &  & RMSE & (1.65) & (0.5) & (0.5) & (0.28) & (0.15) &  & (4.65) & (5.09) & (5.17) & (4.69) & (5.12) \\
 & T & Bias & 0.46 & 0.17 & 0.15 & 0.05 & 0.01 &  & 3.65 & 4.09 & 4.3 & 3.51 & 4.19 \\
 &  & RMSE & (1.69) & (0.5) & (0.5) & (0.28) & (0.16) &  & (4.64) & (5.09) & (5.17) & (4.72) & (5.14) \\
 & U & Bias & 0.46 & 0.17 & 0.15 & 0.05 & 0.01 &  & 3.63 & 4.12 & 4.43 & 3.56 & 4.23 \\
 &  & RMSE & (1.69) & (0.5) & (0.5) & (0.29) & (0.16) &  & (4.62) & (5.1) & (5.25) & (4.73) & (5.14) \\
 & Non & Bias & 0.46 & 0.17 & 0.15 & 0.05 & 0.01 &  & 3.63 & 4.12 & 4.43 & 3.56 & 4.23 \\
 &  & RMSE & (1.69) & (0.5) & (0.5) & (0.29) & (0.16) &  & (4.62) & (5.1) & (5.25) & (4.73) & (5.14) \\
 &  &  &  &  &  &  &  &  &  &  &  &  &  \\
0.5 & unmatched & Mean Num & 0.01 & 0.0 & 0.0 & 0.0 & 0.02 & 0.0 & 0.01 & 0.0 & 0.0 & 0.0 & 0.02 \\
 & U,T & Bias & 0.62 & 0.13 & 0.15 & 0.05 & 0.01 &  & 4.56 & 4.87 & 5.3 & 4.39 & 5.14 \\
 &  & RMSE & (1.99) & (0.44) & (0.5) & (0.28) & (0.15) &  & (5.52) & (5.74) & (6.03) & (5.4) & (5.93) \\
 & T & Bias & 0.62 & 0.13 & 0.15 & 0.05 & 0.01 &  & 4.53 & 4.87 & 5.3 & 4.46 & 5.16 \\
 &  & RMSE & (1.99) & (0.44) & (0.5) & (0.28) & (0.16) &  & (5.5) & (5.74) & (6.03) & (5.45) & (5.95) \\
 & U & Bias & 0.62 & 0.13 & 0.15 & 0.05 & 0.01 &  & 4.53 & 4.9 & 5.43 & 4.56 & 5.22 \\
 &  & RMSE & (1.99) & (0.44) & (0.5) & (0.29) & (0.16) &  & (5.5) & (5.75) & (6.12) & (5.52) & (5.98) \\
 & Non & Bias & 0.62 & 0.13 & 0.15 & 0.05 & 0.01 &  & 4.53 & 4.9 & 5.43 & 4.56 & 5.22 \\
 &  & RMSE & (1.99) & (0.44) & (0.5) & (0.29) & (0.16) &  & (5.5) & (5.75) & (6.12) & (5.52) & (5.98) \\
 &  &  &  &  &  &  &  &  &  &  &  &  &  \\
0.5 & unmatched & Mean Num & 1.6 & 2.83 & 3.91 & 5.74 & 10.73 & -1.0 & 1.6 & 2.83 & 3.91 & 5.74 & 10.73 \\
 & U,T & Bias & 0.49 & 0.41 & 0.21 & 0.12 & 0.07 &  & 1.98 & 0.6 & 0.34 & 0.17 & -0.07 \\
 &  & RMSE & (1.54) & (1.13) & (0.52) & (0.3) & (0.23) &  & (4.4) & (2.75) & (1.75) & (1.22) & (0.31) \\
 & T & Bias & 0.6 & 0.38 & 0.21 & 0.11 & 0.07 &  & 2.18 & 2.15 & 1.86 & 2.07 & 1.81 \\
 &  & RMSE & (1.77) & (1.17) & (0.51) & (0.32) & (0.22) &  & (4.34) & (4.27) & (3.59) & (4.05) & (3.81) \\
 & U & Bias & 0.49 & 0.41 & 0.21 & 0.13 & 0.07 &  & 1.95 & 0.52 & 0.35 & 0.12 & -0.04 \\
 &  & RMSE & (1.54) & (1.13) & (0.52) & (0.31) & (0.22) &  & (4.35) & (2.64) & (1.75) & (1.16) & (0.27) \\
 & Non & Bias & 0.6 & 0.37 & 0.21 & 0.12 & 0.07 &  & 2.15 & 2.2 & 1.82 & 1.95 & 1.44 \\
 &  & RMSE & (1.77) & (1.17) & (0.51) & (0.33) & (0.22) &  & (4.3) & (4.34) & (3.56) & (3.91) & (3.3) \\
 &  &  &  &  &  &  &  &  &  &  &  &  &  \\
0.5 & unmatched & Mean Num & 5.66 & 11.08 & 16.08 & 26.43 & 50.88 & -2.0 & 5.66 & 11.08 & 16.08 & 26.43 & 50.88 \\
 & U,T & Bias & 0.35 & 0.16 & 0.14 & 0.1 & 0.04 &  & -0.59 & -0.53 & -0.41 & -0.3 & -0.59 \\
 &  & RMSE & (0.97) & (0.4) & (0.32) & (0.23) & (0.14) &  & (1.46) & (1.42) & (1.08) & (0.61) & (1.48) \\
 & T & Bias & 0.66 & 0.67 & 0.35 & 0.42 & 0.09 &  & 1.83 & 3.45 & 4.74 & 5.11 & 5.31 \\
 &  & RMSE & (1.8) & (2.18) & (1.51) & (1.06) & (0.31) &  & (4.21) & (5.55) & (6.27) & (6.45) & (6.6) \\
 & U & Bias & 0.35 & 0.15 & 0.13 & 0.08 & 0.03 &  & -0.58 & -0.52 & -0.39 & -0.28 & -0.61 \\
 &  & RMSE & (0.98) & (0.38) & (0.3) & (0.2) & (0.14) &  & (1.45) & (1.41) & (1.07) & (0.59) & (1.51) \\
 & Non & Bias & 0.68 & 0.67 & 0.4 & 0.42 & 0.09 &  & 1.9 & 3.38 & 4.81 & 5.09 & 5.43 \\
 &  & RMSE & (1.8) & (2.18) & (1.58) & (1.06) & (0.32) &  & (4.27) & (5.5) & (6.35) & (6.41) & (6.71) \\
 &  &  &  &  &  &  &  &  &  &  &  &  &  \\
0.5 & unmatched & Mean Num & 8.85 & 17.29 & 25.92 & 42.62 & 84.41 & -3.0 & 8.85 & 17.29 & 25.92 & 42.62 & 84.41 \\
 & U,T & Bias & -0.24 & 0.28 & 0.14 & 0.03 & 0.03 &  & -1.57 & -1.32 & -0.82 & -1.0 & -0.99 \\
 &  & RMSE & (2.41) & (0.83) & (0.46) & (0.19) & (0.17) &  & (2.86) & (2.47) & (1.74) & (2.09) & (2.13) \\
 & T & Bias & 1.41 & 1.49 & 1.85 & 0.84 & 1.36 &  & 1.49 & 4.68 & 5.08 & 6.1 & 5.88 \\
 &  & RMSE & (4.14) & (3.58) & (3.99) & (2.47) & (2.95) &  & (5.47) & (6.86) & (6.96) & (7.43) & (7.09) \\
 & U & Bias & -0.25 & 0.27 & 0.13 & 0.04 & 0.02 &  & -1.56 & -1.29 & -0.8 & -1.07 & -1.04 \\
 &  & RMSE & (2.41) & (0.82) & (0.44) & (0.19) & (0.14) &  & (2.86) & (2.46) & (1.71) & (2.16) & (2.15) \\
 & Non & Bias & 1.45 & 1.49 & 1.84 & 0.83 & 1.37 &  & 1.52 & 4.68 & 5.06 & 6.1 & 5.85 \\
 &  & RMSE & (4.17) & (3.58) & (3.99) & (2.47) & (2.95) &  & (5.48) & (6.86) & (6.94) & (7.43) & (7.07) \\
 &  &  &  &  &  &  &  &  &  &  &  &  &  \\
\hline 
\bottomrule 
\end{tabular}

\begin{tablenotes}
\item[a]\textit{Note:}The objective function was numerically maximized using differential evolution (DE) algorithm in \texttt{BlackBoxOptim.jl} package. For DE algorithm, I require setting the domain of parameters and the number of population seeds so that I fix the former to $[-10, 10]$ for parameters and the latter to 400. The reported point estimates are the best-found maxima. I omit the parameter $\beta_0$ that can take on only a finite number of values (here +1) converge at an arbitrarily fast rate, they are superconsistent. 
\end{tablenotes}
}\end{center}
\label{tb:estimation_results_single_market_two_param_beta_with_dummy_with_IR_penalty_1}
\end{table}

\begin{table}[!ht]
\caption{\textbf{Case 2 with IR ($\lambda = 2$).}}
\begin{center}
{\tiny
  \begin{tabular}{@{\extracolsep{5pt}}lc|cccccc|lccccc}
\toprule 
 & Num of agents &  & 10 & 20 & 30 & 50 & 100 &  & 10 & 20 & 30 & 50 & 100 \\
$\beta_1$ &  &  &  &  &  &  &  & $\beta_2$ &  &  &  &  &  \\
\midrule 
0.5 & unmatched & Mean Num & 0.0 & 0.0 & 0.0 & 0.0 & 0.0 & 1.0 & 0.0 & 0.0 & 0.0 & 0.0 & 0.0 \\
 & U,T & Bias & 0.04 & 0.09 & 0.14 & 0.06 & 0.02 &  & 4.3 & 4.04 & 3.96 & 3.7 & 4.45 \\
 &  & RMSE & (0.22) & (0.31) & (0.48) & (0.29) & (0.16) &  & (5.13) & (4.96) & (4.93) & (4.59) & (5.44) \\
 & T & Bias & 0.04 & 0.08 & 0.14 & 0.06 & 0.01 &  & 4.3 & 4.03 & 3.9 & 3.59 & 4.46 \\
 &  & RMSE & (0.22) & (0.29) & (0.48) & (0.29) & (0.17) &  & (5.13) & (4.94) & (4.86) & (4.48) & (5.45) \\
 & U & Bias & 0.07 & 0.09 & 0.14 & 0.06 & 0.01 &  & 4.3 & 4.04 & 3.88 & 3.64 & 4.46 \\
 &  & RMSE & (0.41) & (0.31) & (0.48) & (0.29) & (0.17) &  & (5.12) & (4.96) & (4.85) & (4.54) & (5.44) \\
 & Non & Bias & 0.04 & 0.08 & 0.14 & 0.06 & 0.01 &  & 4.33 & 4.04 & 3.93 & 3.57 & 4.46 \\
 &  & RMSE & (0.22) & (0.29) & (0.48) & (0.29) & (0.17) &  & (5.13) & (4.95) & (4.89) & (4.47) & (5.44) \\
 &  &  &  &  &  &  &  &  &  &  &  &  &  \\
0.5 & unmatched & Mean Num & 0.02 & 0.0 & 0.0 & 0.0 & 0.02 & 0.0 & 0.02 & 0.0 & 0.0 & 0.0 & 0.02 \\
 & U,T & Bias & 0.23 & 0.07 & 0.14 & 0.05 & 0.01 &  & 5.15 & 5.19 & 4.95 & 4.82 & 5.42 \\
 &  & RMSE & (1.3) & (0.25) & (0.48) & (0.29) & (0.16) &  & (5.91) & (5.97) & (5.72) & (5.58) & (6.25) \\
 & T & Bias & 0.23 & 0.08 & 0.14 & 0.06 & 0.02 &  & 5.15 & 5.21 & 4.99 & 4.7 & 5.42 \\
 &  & RMSE & (1.3) & (0.27) & (0.48) & (0.29) & (0.16) &  & (5.91) & (6.0) & (5.75) & (5.45) & (6.24) \\
 & U & Bias & 0.26 & 0.07 & 0.14 & 0.06 & 0.01 &  & 5.16 & 5.2 & 4.97 & 4.71 & 5.42 \\
 &  & RMSE & (1.34) & (0.25) & (0.48) & (0.29) & (0.17) &  & (5.91) & (5.98) & (5.74) & (5.46) & (6.25) \\
 & Non & Bias & 0.26 & 0.08 & 0.14 & 0.06 & 0.01 &  & 5.16 & 5.22 & 5.03 & 4.64 & 5.42 \\
 &  & RMSE & (1.34) & (0.27) & (0.48) & (0.29) & (0.17) &  & (5.91) & (6.01) & (5.78) & (5.38) & (6.25) \\
 &  &  &  &  &  &  &  &  &  &  &  &  &  \\
0.5 & unmatched & Mean Num & 1.53 & 2.87 & 3.9 & 5.85 & 10.63 & -1.0 & 1.53 & 2.87 & 3.9 & 5.85 & 10.63 \\
 & U,T & Bias & 0.52 & 0.33 & 0.22 & 0.1 & 0.06 &  & 1.86 & 0.75 & 1.36 & 0.56 & 0.03 \\
 &  & RMSE & (1.56) & (1.03) & (0.5) & (0.3) & (0.21) &  & (4.44) & (2.75) & (3.59) & (1.93) & (0.28) \\
 & T & Bias & 0.54 & 0.25 & 0.16 & 0.1 & 0.06 &  & 2.21 & 2.05 & 2.24 & 2.09 & 1.58 \\
 &  & RMSE & (1.57) & (0.96) & (0.44) & (0.31) & (0.21) &  & (4.74) & (4.12) & (4.19) & (3.82) & (3.32) \\
 & U & Bias & 0.53 & 0.33 & 0.21 & 0.11 & 0.06 &  & 1.83 & 0.8 & 1.2 & 0.55 & 0.04 \\
 &  & RMSE & (1.56) & (1.03) & (0.5) & (0.31) & (0.22) &  & (4.4) & (2.85) & (3.36) & (1.93) & (0.27) \\
 & Non & Bias & 0.47 & 0.23 & 0.16 & 0.11 & 0.06 &  & 2.39 & 2.16 & 2.24 & 1.99 & 1.55 \\
 &  & RMSE & (1.48) & (0.96) & (0.44) & (0.32) & (0.21) &  & (4.83) & (4.21) & (4.19) & (3.73) & (3.36) \\
 &  &  &  &  &  &  &  &  &  &  &  &  &  \\
0.5 & unmatched & Mean Num & 5.64 & 11.15 & 16.06 & 26.17 & 50.89 & -2.0 & 5.64 & 11.15 & 16.06 & 26.17 & 50.89 \\
 & U,T & Bias & 0.36 & 0.18 & 0.16 & 0.1 & 0.07 &  & -0.57 & -0.32 & -0.2 & -0.15 & -0.22 \\
 &  & RMSE & (0.98) & (0.42) & (0.33) & (0.28) & (0.19) &  & (1.42) & (0.85) & (0.52) & (0.45) & (0.57) \\
 & T & Bias & 0.42 & 0.1 & 0.08 & 0.21 & 0.05 &  & 1.82 & 4.31 & 4.51 & 5.35 & 5.64 \\
 &  & RMSE & (1.5) & (0.61) & (0.22) & (0.65) & (0.23) &  & (4.23) & (6.08) & (6.02) & (6.58) & (6.88) \\
 & U & Bias & 0.36 & 0.17 & 0.15 & 0.08 & 0.07 &  & -0.56 & -0.3 & -0.18 & -0.13 & -0.26 \\
 &  & RMSE & (0.98) & (0.41) & (0.32) & (0.25) & (0.2) &  & (1.42) & (0.85) & (0.49) & (0.4) & (0.65) \\
 & Non & Bias & 0.47 & 0.12 & 0.13 & 0.21 & 0.05 &  & 1.92 & 4.19 & 4.49 & 5.31 & 5.57 \\
 &  & RMSE & (1.78) & (0.63) & (0.54) & (0.65) & (0.23) &  & (4.39) & (6.01) & (5.98) & (6.53) & (6.82) \\
 &  &  &  &  &  &  &  &  &  &  &  &  &  \\
0.5 & unmatched & Mean Num & 8.81 & 17.34 & 25.76 & 42.53 & 84.41 & -3.0 & 8.81 & 17.34 & 25.76 & 42.53 & 84.41 \\
 & U,T & Bias & -0.26 & 0.31 & 0.18 & 0.08 & 0.07 &  & -1.58 & -1.13 & -0.63 & -0.41 & -0.51 \\
 &  & RMSE & (2.36) & (0.82) & (0.44) & (0.24) & (0.19) &  & (2.82) & (2.19) & (1.4) & (1.04) & (1.38) \\
 & T & Bias & 1.29 & 0.89 & 0.37 & 0.11 & 0.08 &  & 1.4 & 5.0 & 5.23 & 5.97 & 6.1 \\
 &  & RMSE & (3.94) & (3.01) & (1.9) & (0.98) & (0.32) &  & (5.39) & (7.26) & (6.77) & (7.09) & (7.32) \\
 & U & Bias & -0.25 & 0.29 & 0.17 & 0.08 & 0.06 &  & -1.58 & -1.1 & -0.62 & -0.45 & -0.55 \\
 &  & RMSE & (2.36) & (0.8) & (0.44) & (0.24) & (0.18) &  & (2.82) & (2.17) & (1.39) & (1.08) & (1.41) \\
 & Non & Bias & 1.32 & 1.08 & 0.35 & 0.19 & 0.1 &  & 1.53 & 5.05 & 5.23 & 5.96 & 6.18 \\
 &  & RMSE & (3.91) & (3.18) & (1.89) & (1.34) & (0.37) &  & (5.44) & (7.39) & (6.72) & (7.11) & (7.38) \\
 &  &  &  &  &  &  &  &  &  &  &  &  &  \\
\hline 
\bottomrule 
\end{tabular}

\begin{tablenotes}
\item[a]\textit{Note:}The objective function was numerically maximized using differential evolution (DE) algorithm in \texttt{BlackBoxOptim.jl} package. For DE algorithm, I require setting the domain of parameters and the number of population seeds so that I fix the former to $[-10, 10]$ for parameters and the latter to 400. The reported point estimates are the best-found maxima. I omit the parameter $\beta_0$ that can take on only a finite number of values (here +1) converge at an arbitrarily fast rate, they are superconsistent. 
\end{tablenotes}
}\end{center}
\label{tb:estimation_results_single_market_two_param_beta_with_dummy_with_IR_penalty_2}
\end{table}

\begin{table}[!ht]
\caption{\textbf{Case 2 with IR ($\lambda = 5$).}}
\begin{center}
{\tiny
  \begin{tabular}{@{\extracolsep{5pt}}lc|cccccc|lccccc}
\toprule 
 & Num of agents &  & 10 & 20 & 30 & 50 & 100 &  & 10 & 20 & 30 & 50 & 100 \\
$\beta_1$ &  &  &  &  &  &  &  & $\beta_2$ &  &  &  &  &  \\
\midrule 
0.5 & unmatched & Mean Num & 0.0 & 0.0 & 0.0 & 0.0 & 0.0 & 1.0 & 0.0 & 0.0 & 0.0 & 0.0 & 0.0 \\
 & U,T & Bias & 0.05 & 0.09 & 0.14 & 0.04 & 0.01 &  & 3.79 & 4.15 & 4.17 & 4.42 & 4.12 \\
 &  & RMSE & (0.18) & (0.28) & (0.44) & (0.27) & (0.16) &  & (4.73) & (5.01) & (5.15) & (5.27) & (4.96) \\
 & T & Bias & 0.05 & 0.08 & 0.14 & 0.04 & 0.0 &  & 3.79 & 4.28 & 4.18 & 4.43 & 4.1 \\
 &  & RMSE & (0.18) & (0.26) & (0.44) & (0.27) & (0.16) &  & (4.73) & (5.06) & (5.15) & (5.28) & (4.95) \\
 & U & Bias & 0.08 & 0.08 & 0.14 & 0.04 & 0.0 &  & 3.78 & 4.22 & 4.14 & 4.48 & 4.11 \\
 &  & RMSE & (0.39) & (0.26) & (0.44) & (0.27) & (0.16) &  & (4.65) & (5.04) & (5.13) & (5.33) & (4.95) \\
 & Non & Bias & 0.08 & 0.08 & 0.14 & 0.05 & 0.0 &  & 3.78 & 4.3 & 4.16 & 4.44 & 4.1 \\
 &  & RMSE & (0.39) & (0.26) & (0.44) & (0.28) & (0.16) &  & (4.65) & (5.09) & (5.14) & (5.29) & (4.97) \\
 &  &  &  &  &  &  &  &  &  &  &  &  &  \\
0.5 & unmatched & Mean Num & 0.01 & 0.0 & 0.0 & 0.0 & 0.02 & 0.0 & 0.01 & 0.0 & 0.0 & 0.0 & 0.02 \\
 & U,T & Bias & 0.04 & 0.07 & 0.06 & 0.04 & 0.01 &  & 4.94 & 5.1 & 5.15 & 5.4 & 4.99 \\
 &  & RMSE & (0.2) & (0.23) & (0.33) & (0.27) & (0.16) &  & (5.69) & (5.72) & (5.91) & (6.14) & (5.7) \\
 & T & Bias & 0.04 & 0.07 & 0.06 & 0.04 & 0.01 &  & 4.94 & 5.01 & 5.15 & 5.43 & 5.07 \\
 &  & RMSE & (0.2) & (0.24) & (0.33) & (0.27) & (0.16) &  & (5.69) & (5.66) & (5.91) & (6.15) & (5.78) \\
 & U & Bias & 0.07 & 0.07 & 0.05 & 0.04 & 0.0 &  & 4.92 & 5.12 & 5.14 & 5.41 & 5.07 \\
 &  & RMSE & (0.4) & (0.23) & (0.34) & (0.27) & (0.16) &  & (5.62) & (5.74) & (5.9) & (6.15) & (5.77) \\
 & Non & Bias & 0.04 & 0.07 & 0.05 & 0.05 & 0.0 &  & 4.99 & 5.03 & 5.13 & 5.44 & 5.1 \\
 &  & RMSE & (0.2) & (0.24) & (0.34) & (0.28) & (0.16) &  & (5.71) & (5.68) & (5.89) & (6.16) & (5.82) \\
 &  &  &  &  &  &  &  &  &  &  &  &  &  \\
0.5 & unmatched & Mean Num & 1.68 & 2.9 & 3.85 & 5.83 & 10.72 & -1.0 & 1.68 & 2.9 & 3.85 & 5.83 & 10.72 \\
 & U,T & Bias & 0.24 & 0.4 & 0.21 & 0.18 & 0.16 &  & 3.13 & 1.4 & 0.92 & 0.82 & 0.35 \\
 &  & RMSE & (0.6) & (1.16) & (0.49) & (0.42) & (0.3) &  & (4.85) & (3.41) & (2.31) & (2.36) & (1.59) \\
 & T & Bias & 0.23 & 0.33 & 0.18 & 0.15 & 0.11 &  & 3.59 & 2.81 & 2.38 & 2.47 & 1.96 \\
 &  & RMSE & (0.58) & (1.09) & (0.47) & (0.4) & (0.26) &  & (5.25) & (4.85) & (4.22) & (4.17) & (3.78) \\
 & U & Bias & 0.24 & 0.4 & 0.2 & 0.19 & 0.16 &  & 3.13 & 1.36 & 0.85 & 0.81 & 0.36 \\
 &  & RMSE & (0.6) & (1.16) & (0.49) & (0.43) & (0.31) &  & (4.85) & (3.38) & (2.21) & (2.36) & (1.59) \\
 & Non & Bias & 0.24 & 0.32 & 0.18 & 0.15 & 0.12 &  & 3.72 & 2.75 & 2.38 & 2.43 & 1.82 \\
 &  & RMSE & (0.59) & (1.09) & (0.47) & (0.4) & (0.27) &  & (5.36) & (4.8) & (4.22) & (4.15) & (3.57) \\
 &  &  &  &  &  &  &  &  &  &  &  &  &  \\
0.5 & unmatched & Mean Num & 5.6 & 11.14 & 16.01 & 26.32 & 50.83 & -2.0 & 5.6 & 11.14 & 16.01 & 26.32 & 50.83 \\
 & U,T & Bias & 0.48 & 0.17 & 0.17 & 0.14 & 0.12 &  & -0.56 & -0.1 & -0.1 & -0.09 & -0.14 \\
 &  & RMSE & (1.2) & (0.45) & (0.35) & (0.3) & (0.23) &  & (1.56) & (0.61) & (0.46) & (0.39) & (0.45) \\
 & T & Bias & 0.49 & 0.1 & 0.05 & 0.08 & 0.08 &  & 1.76 & 4.73 & 4.47 & 5.65 & 5.71 \\
 &  & RMSE & (1.58) & (0.64) & (0.25) & (0.24) & (0.31) &  & (4.08) & (6.39) & (5.95) & (6.97) & (6.69) \\
 & U & Bias & 0.48 & 0.16 & 0.16 & 0.15 & 0.12 &  & -0.56 & -0.09 & -0.08 & -0.1 & -0.15 \\
 &  & RMSE & (1.2) & (0.43) & (0.33) & (0.3) & (0.24) &  & (1.56) & (0.59) & (0.45) & (0.39) & (0.44) \\
 & Non & Bias & 0.57 & 0.1 & 0.1 & 0.08 & 0.08 &  & 1.88 & 4.72 & 4.53 & 5.65 & 5.69 \\
 &  & RMSE & (1.8) & (0.65) & (0.55) & (0.24) & (0.31) &  & (4.22) & (6.41) & (6.02) & (6.97) & (6.68) \\
 &  &  &  &  &  &  &  &  &  &  &  &  &  \\
0.5 & unmatched & Mean Num & 8.93 & 17.36 & 25.72 & 42.63 & 84.3 & -3.0 & 8.93 & 17.36 & 25.72 & 42.63 & 84.3 \\
 & U,T & Bias & -0.26 & 0.3 & 0.17 & 0.12 & 0.07 &  & -1.44 & -0.84 & -0.47 & -0.25 & -0.16 \\
 &  & RMSE & (2.4) & (0.84) & (0.5) & (0.28) & (0.18) &  & (2.77) & (1.85) & (1.37) & (0.84) & (0.71) \\
 & T & Bias & 1.27 & 0.93 & 0.38 & 0.11 & 0.08 &  & 1.45 & 5.17 & 5.36 & 5.94 & 6.43 \\
 &  & RMSE & (3.95) & (3.08) & (1.9) & (0.98) & (0.33) &  & (5.37) & (7.15) & (6.92) & (6.96) & (7.43) \\
 & U & Bias & -0.26 & 0.3 & 0.18 & 0.12 & 0.08 &  & -1.45 & -0.83 & -0.48 & -0.25 & -0.18 \\
 &  & RMSE & (2.4) & (0.84) & (0.5) & (0.28) & (0.19) &  & (2.78) & (1.85) & (1.37) & (0.84) & (0.72) \\
 & Non & Bias & 1.3 & 1.04 & 0.37 & 0.2 & 0.1 &  & 1.58 & 5.28 & 5.39 & 6.0 & 6.4 \\
 &  & RMSE & (3.92) & (3.16) & (1.9) & (1.34) & (0.38) &  & (5.43) & (7.25) & (6.89) & (7.06) & (7.41) \\
 &  &  &  &  &  &  &  &  &  &  &  &  &  \\
\hline 
\bottomrule 
\end{tabular}

\begin{tablenotes}
\item[a]\textit{Note:}The objective function was numerically maximized using differential evolution (DE) algorithm in \texttt{BlackBoxOptim.jl} package. For DE algorithm, I require setting the domain of parameters and the number of population seeds so that I fix the former to $[-10, 10]$ for parameters and the latter to 400. The reported point estimates are the best-found maxima. I omit the parameter $\beta_0$ that can take on only a finite number of values (here +1) converge at an arbitrarily fast rate, they are superconsistent. 
\end{tablenotes}
}\end{center}
\label{tb:estimation_results_single_market_two_param_beta_with_dummy_with_IR_penalty_5}
\end{table}

\begin{table}[!ht]
\caption{\textbf{Case 2 with IR ($\lambda = 10$).}}
\begin{center}
{\tiny
  \begin{tabular}{@{\extracolsep{5pt}}lc|cccccc|lccccc}
\toprule 
 & Num of agents &  & 10 & 20 & 30 & 50 & 100 &  & 10 & 20 & 30 & 50 & 100 \\
$\beta_1$ &  &  &  &  &  &  &  & $\beta_2$ &  &  &  &  &  \\
\midrule 
0.5 & unmatched & Mean Num & 0.0 & 0.0 & 0.0 & 0.0 & 0.0 & 1.0 & 0.0 & 0.0 & 0.0 & 0.0 & 0.0 \\
 & U,T & Bias & 0.04 & 0.13 & 0.16 & 0.05 & 0.0 &  & 3.61 & 4.73 & 4.25 & 4.16 & 4.49 \\
 &  & RMSE & (0.18) & (0.53) & (0.48) & (0.25) & (0.16) &  & (4.68) & (5.4) & (5.09) & (5.01) & (5.34) \\
 & T & Bias & 0.04 & 0.13 & 0.16 & 0.05 & 0.01 &  & 3.61 & 4.73 & 4.27 & 4.24 & 4.5 \\
 &  & RMSE & (0.18) & (0.53) & (0.48) & (0.26) & (0.16) &  & (4.68) & (5.4) & (5.12) & (5.09) & (5.34) \\
 & U & Bias & 0.07 & 0.13 & 0.15 & 0.06 & 0.01 &  & 3.63 & 4.66 & 4.3 & 4.2 & 4.52 \\
 &  & RMSE & (0.39) & (0.53) & (0.48) & (0.26) & (0.16) &  & (4.65) & (5.35) & (5.13) & (5.05) & (5.36) \\
 & Non & Bias & 0.07 & 0.13 & 0.15 & 0.06 & 0.01 &  & 3.63 & 4.66 & 4.22 & 4.2 & 4.52 \\
 &  & RMSE & (0.39) & (0.53) & (0.48) & (0.26) & (0.16) &  & (4.65) & (5.35) & (5.07) & (5.05) & (5.36) \\
 &  &  &  &  &  &  &  &  &  &  &  &  &  \\
0.5 & unmatched & Mean Num & 0.01 & 0.0 & 0.0 & 0.0 & 0.02 & 0.0 & 0.01 & 0.0 & 0.0 & 0.0 & 0.02 \\
 & U,T & Bias & 0.05 & 0.1 & 0.16 & 0.06 & 0.01 &  & 4.84 & 5.69 & 5.28 & 5.07 & 5.61 \\
 &  & RMSE & (0.19) & (0.37) & (0.48) & (0.28) & (0.16) &  & (5.69) & (6.32) & (5.97) & (5.86) & (6.28) \\
 & T & Bias & 0.05 & 0.1 & 0.16 & 0.06 & 0.01 &  & 4.84 & 5.69 & 5.28 & 5.09 & 5.61 \\
 &  & RMSE & (0.19) & (0.37) & (0.48) & (0.28) & (0.16) &  & (5.69) & (6.32) & (5.97) & (5.85) & (6.28) \\
 & U & Bias & 0.05 & 0.1 & 0.15 & 0.06 & 0.01 &  & 4.88 & 5.62 & 5.22 & 5.08 & 5.6 \\
 &  & RMSE & (0.19) & (0.37) & (0.48) & (0.28) & (0.16) &  & (5.7) & (6.27) & (5.93) & (5.86) & (6.29) \\
 & Non & Bias & 0.09 & 0.1 & 0.15 & 0.06 & 0.01 &  & 4.88 & 5.62 & 5.31 & 5.05 & 5.6 \\
 &  & RMSE & (0.39) & (0.37) & (0.48) & (0.29) & (0.16) &  & (5.7) & (6.27) & (5.99) & (5.82) & (6.29) \\
 &  &  &  &  &  &  &  &  &  &  &  &  &  \\
0.5 & unmatched & Mean Num & 1.44 & 2.89 & 3.84 & 5.72 & 10.67 & -1.0 & 1.44 & 2.89 & 3.84 & 5.72 & 10.67 \\
 & U,T & Bias & 0.48 & 0.3 & 0.21 & 0.19 & 0.14 &  & 3.64 & 2.26 & 1.52 & 1.1 & 1.17 \\
 &  & RMSE & (1.49) & (0.69) & (0.53) & (0.41) & (0.32) &  & (5.6) & (4.1) & (3.22) & (2.56) & (2.82) \\
 & T & Bias & 0.47 & 0.24 & 0.22 & 0.2 & 0.11 &  & 3.8 & 3.28 & 2.74 & 2.51 & 2.46 \\
 &  & RMSE & (1.49) & (0.59) & (0.54) & (0.44) & (0.29) &  & (5.72) & (4.9) & (4.43) & (4.26) & (4.16) \\
 & U & Bias & 0.48 & 0.3 & 0.2 & 0.2 & 0.14 &  & 3.64 & 2.18 & 1.47 & 1.12 & 1.18 \\
 &  & RMSE & (1.49) & (0.7) & (0.53) & (0.42) & (0.32) &  & (5.6) & (4.02) & (3.19) & (2.63) & (2.82) \\
 & Non & Bias & 0.4 & 0.22 & 0.22 & 0.2 & 0.11 &  & 3.96 & 3.26 & 2.66 & 2.45 & 2.55 \\
 &  & RMSE & (1.22) & (0.59) & (0.54) & (0.44) & (0.29) &  & (5.78) & (4.88) & (4.34) & (4.11) & (4.26) \\
 &  &  &  &  &  &  &  &  &  &  &  &  &  \\
0.5 & unmatched & Mean Num & 5.54 & 11.09 & 15.98 & 26.51 & 51.44 & -2.0 & 5.54 & 11.09 & 15.98 & 26.51 & 51.44 \\
 & U,T & Bias & 0.65 & 0.24 & 0.2 & 0.23 & 0.15 &  & -0.68 & -0.11 & -0.06 & -0.13 & -0.07 \\
 &  & RMSE & (1.22) & (0.52) & (0.45) & (0.41) & (0.3) &  & (1.61) & (0.7) & (0.55) & (0.51) & (0.39) \\
 & T & Bias & 0.55 & 0.11 & 0.2 & 0.19 & 0.13 &  & 1.91 & 4.6 & 4.79 & 5.31 & 5.74 \\
 &  & RMSE & (1.51) & (0.66) & (0.97) & (0.51) & (0.35) &  & (4.24) & (6.21) & (6.16) & (6.64) & (7.03) \\
 & U & Bias & 0.65 & 0.25 & 0.2 & 0.21 & 0.15 &  & -0.68 & -0.13 & -0.07 & -0.11 & -0.06 \\
 &  & RMSE & (1.22) & (0.57) & (0.45) & (0.39) & (0.3) &  & (1.61) & (0.74) & (0.56) & (0.49) & (0.38) \\
 & Non & Bias & 0.61 & 0.12 & 0.25 & 0.19 & 0.13 &  & 1.86 & 4.65 & 4.85 & 5.33 & 5.77 \\
 &  & RMSE & (1.77) & (0.68) & (1.09) & (0.51) & (0.35) &  & (4.22) & (6.23) & (6.24) & (6.65) & (7.06) \\
 &  &  &  &  &  &  &  &  &  &  &  &  &  \\
0.5 & unmatched & Mean Num & 8.76 & 17.34 & 25.8 & 42.51 & 84.2 & -3.0 & 8.76 & 17.34 & 25.8 & 42.51 & 84.2 \\
 & U,T & Bias & -0.23 & 0.37 & 0.24 & 0.14 & 0.09 &  & -1.54 & -0.76 & -0.41 & -0.12 & -0.07 \\
 &  & RMSE & (2.39) & (0.79) & (0.59) & (0.3) & (0.25) &  & (2.85) & (1.7) & (1.27) & (0.57) & (0.51) \\
 & T & Bias & 1.29 & 1.01 & 0.42 & 0.1 & 0.08 &  & 1.41 & 5.31 & 5.42 & 6.19 & 6.76 \\
 &  & RMSE & (3.94) & (3.22) & (1.91) & (0.99) & (0.34) &  & (5.41) & (7.22) & (7.04) & (7.16) & (7.72) \\
 & U & Bias & -0.23 & 0.36 & 0.2 & 0.14 & 0.08 &  & -1.54 & -0.73 & -0.33 & -0.13 & -0.04 \\
 &  & RMSE & (2.39) & (0.79) & (0.52) & (0.3) & (0.22) &  & (2.85) & (1.69) & (1.08) & (0.56) & (0.46) \\
 & Non & Bias & 1.34 & 1.05 & 0.4 & 0.1 & 0.11 &  & 1.52 & 5.33 & 5.48 & 6.18 & 6.75 \\
 &  & RMSE & (3.92) & (3.22) & (1.9) & (0.99) & (0.42) &  & (5.47) & (7.22) & (7.01) & (7.15) & (7.72) \\
 &  &  &  &  &  &  &  &  &  &  &  &  &  \\
\hline 
\bottomrule 
\end{tabular}

\begin{tablenotes}
\item[a]\textit{Note:}The objective function was numerically maximized using differential evolution (DE) algorithm in \texttt{BlackBoxOptim.jl} package. For DE algorithm, I require setting the domain of parameters and the number of population seeds so that I fix the former to $[-10, 10]$ for parameters and the latter to 400. The reported point estimates are the best-found maxima. I omit the parameter $\beta_0$ that can take on only a finite number of values (here +1) converge at an arbitrarily fast rate, they are superconsistent. 
\end{tablenotes}
}\end{center}
\label{tb:estimation_results_single_market_two_param_beta_with_dummy_with_IR_penalty_10}
\end{table}

\begin{table}[!ht]
\caption{\textbf{Case 2 with IR ($\lambda = 20$).}}
\begin{center}
{\tiny
  \begin{tabular}{@{\extracolsep{5pt}}lc|cccccc|lccccc}
\toprule 
 & Num of agents &  & 10 & 20 & 30 & 50 & 100 &  & 10 & 20 & 30 & 50 & 100 \\
$\beta_1$ &  &  &  &  &  &  &  & $\beta_2$ &  &  &  &  &  \\
\midrule 
0.5 & unmatched & Mean Num & 0.0 & 0.0 & 0.0 & 0.0 & 0.0 & 1.0 & 0.0 & 0.0 & 0.0 & 0.0 & 0.0 \\
 & U,T & Bias & 0.12 & 0.05 & 0.04 & 0.05 & 0.01 &  & 3.82 & 4.07 & 4.7 & 4.44 & 4.54 \\
 &  & RMSE & (0.89) & (0.22) & (0.29) & (0.27) & (0.17) &  & (4.8) & (4.92) & (5.43) & (5.3) & (5.35) \\
 & T & Bias & 0.12 & 0.05 & 0.04 & 0.05 & 0.01 &  & 3.85 & 4.07 & 4.64 & 4.42 & 4.54 \\
 &  & RMSE & (0.89) & (0.22) & (0.29) & (0.27) & (0.17) &  & (4.84) & (4.92) & (5.38) & (5.3) & (5.35) \\
 & U & Bias & 0.2 & 0.05 & 0.03 & 0.05 & 0.01 &  & 3.8 & 4.06 & 4.67 & 4.37 & 4.51 \\
 &  & RMSE & (1.07) & (0.22) & (0.29) & (0.27) & (0.17) &  & (4.81) & (4.89) & (5.39) & (5.24) & (5.32) \\
 & Non & Bias & 0.12 & 0.05 & 0.04 & 0.05 & 0.01 &  & 3.9 & 4.06 & 4.64 & 4.37 & 4.51 \\
 &  & RMSE & (0.89) & (0.22) & (0.29) & (0.27) & (0.17) &  & (4.85) & (4.89) & (5.38) & (5.24) & (5.32) \\
 &  &  &  &  &  &  &  &  &  &  &  &  &  \\
0.5 & unmatched & Mean Num & 0.01 & 0.0 & 0.0 & 0.0 & 0.02 & 0.0 & 0.01 & 0.0 & 0.0 & 0.0 & 0.02 \\
 & U,T & Bias & 0.04 & 0.11 & 0.12 & 0.05 & 0.02 &  & 5.02 & 5.55 & 5.31 & 5.5 & 5.64 \\
 &  & RMSE & (0.18) & (0.4) & (0.43) & (0.27) & (0.17) &  & (5.84) & (6.16) & (6.05) & (6.23) & (6.31) \\
 & T & Bias & 0.04 & 0.11 & 0.12 & 0.05 & 0.02 &  & 5.02 & 5.55 & 5.26 & 5.38 & 5.64 \\
 &  & RMSE & (0.18) & (0.4) & (0.44) & (0.28) & (0.17) &  & (5.84) & (6.16) & (6.0) & (6.11) & (6.31) \\
 & U & Bias & 0.07 & 0.11 & 0.12 & 0.05 & 0.01 &  & 4.97 & 5.53 & 5.28 & 5.33 & 5.6 \\
 &  & RMSE & (0.39) & (0.4) & (0.44) & (0.28) & (0.17) &  & (5.79) & (6.13) & (6.02) & (6.06) & (6.26) \\
 & Non & Bias & 0.07 & 0.11 & 0.12 & 0.05 & 0.01 &  & 5.0 & 5.53 & 5.28 & 5.33 & 5.6 \\
 &  & RMSE & (0.39) & (0.4) & (0.44) & (0.28) & (0.17) &  & (5.83) & (6.13) & (6.02) & (6.06) & (6.26) \\
 &  &  &  &  &  &  &  &  &  &  &  &  &  \\
0.5 & unmatched & Mean Num & 1.63 & 2.83 & 3.69 & 5.7 & 10.62 & -1.0 & 1.63 & 2.83 & 3.69 & 5.7 & 10.62 \\
 & U,T & Bias & 0.58 & 0.32 & 0.17 & 0.18 & 0.21 &  & 3.79 & 1.84 & 1.56 & 1.37 & 1.26 \\
 &  & RMSE & (1.58) & (0.79) & (0.52) & (0.39) & (0.38) &  & (5.58) & (3.67) & (3.06) & (3.01) & (3.04) \\
 & T & Bias & 0.53 & 0.24 & 0.11 & 0.15 & 0.15 &  & 4.08 & 2.82 & 2.65 & 2.39 & 2.61 \\
 &  & RMSE & (1.54) & (0.59) & (0.43) & (0.39) & (0.33) &  & (5.76) & (4.39) & (4.26) & (3.91) & (4.36) \\
 & U & Bias & 0.58 & 0.32 & 0.17 & 0.18 & 0.2 &  & 3.71 & 1.84 & 1.43 & 1.34 & 1.25 \\
 &  & RMSE & (1.58) & (0.79) & (0.52) & (0.39) & (0.38) &  & (5.53) & (3.67) & (2.89) & (3.04) & (3.04) \\
 & Non & Bias & 0.54 & 0.22 & 0.11 & 0.15 & 0.16 &  & 4.26 & 2.81 & 2.57 & 2.4 & 2.61 \\
 &  & RMSE & (1.54) & (0.59) & (0.43) & (0.39) & (0.33) &  & (5.92) & (4.36) & (4.17) & (3.94) & (4.33) \\
 &  &  &  &  &  &  &  &  &  &  &  &  &  \\
0.5 & unmatched & Mean Num & 5.43 & 11.28 & 16.02 & 26.5 & 50.92 & -2.0 & 5.43 & 11.28 & 16.02 & 26.5 & 50.92 \\
 & U,T & Bias & 0.64 & 0.31 & 0.26 & 0.18 & 0.23 &  & -0.53 & -0.12 & -0.05 & -0.01 & -0.09 \\
 &  & RMSE & (1.42) & (0.57) & (0.47) & (0.36) & (0.42) &  & (1.97) & (0.6) & (0.47) & (0.42) & (0.5) \\
 & T & Bias & 0.52 & 0.13 & 0.11 & 0.19 & 0.11 &  & 2.07 & 4.83 & 5.08 & 5.23 & 5.94 \\
 &  & RMSE & (1.57) & (0.69) & (0.28) & (0.56) & (0.3) &  & (4.38) & (6.38) & (6.53) & (6.49) & (7.23) \\
 & U & Bias & 0.63 & 0.3 & 0.26 & 0.18 & 0.22 &  & -0.52 & -0.11 & -0.05 & -0.01 & -0.06 \\
 &  & RMSE & (1.42) & (0.56) & (0.47) & (0.37) & (0.41) &  & (1.97) & (0.59) & (0.47) & (0.42) & (0.47) \\
 & Non & Bias & 0.61 & 0.13 & 0.15 & 0.19 & 0.11 &  & 2.08 & 4.82 & 5.15 & 5.28 & 5.88 \\
 &  & RMSE & (1.85) & (0.69) & (0.57) & (0.56) & (0.3) &  & (4.39) & (6.37) & (6.6) & (6.55) & (7.18) \\
 &  &  &  &  &  &  &  &  &  &  &  &  &  \\
0.5 & unmatched & Mean Num & 8.8 & 17.38 & 25.7 & 42.57 & 84.31 & -3.0 & 8.8 & 17.38 & 25.7 & 42.57 & 84.31 \\
 & U,T & Bias & -0.26 & 0.43 & 0.19 & 0.19 & 0.13 &  & -1.54 & -0.73 & -0.21 & -0.19 & -0.05 \\
 &  & RMSE & (2.43) & (0.91) & (0.52) & (0.37) & (0.33) &  & (2.94) & (1.73) & (0.99) & (0.64) & (0.55) \\
 & T & Bias & 1.35 & 1.02 & 0.39 & 0.13 & 0.15 &  & 1.46 & 5.33 & 5.56 & 6.18 & 6.32 \\
 &  & RMSE & (4.01) & (3.23) & (1.9) & (0.99) & (0.84) &  & (5.45) & (7.22) & (7.07) & (7.17) & (7.38) \\
 & U & Bias & -0.26 & 0.42 & 0.18 & 0.21 & 0.12 &  & -1.54 & -0.72 & -0.2 & -0.24 & -0.03 \\
 &  & RMSE & (2.43) & (0.91) & (0.5) & (0.4) & (0.31) &  & (2.94) & (1.72) & (0.92) & (0.7) & (0.54) \\
 & Non & Bias & 1.34 & 1.14 & 0.38 & 0.12 & 0.17 &  & 1.6 & 5.43 & 5.48 & 6.2 & 6.29 \\
 &  & RMSE & (3.94) & (3.31) & (1.9) & (0.99) & (0.86) &  & (5.47) & (7.33) & (6.97) & (7.18) & (7.36) \\
 &  &  &  &  &  &  &  &  &  &  &  &  &  \\
\hline 
\bottomrule 
\end{tabular}

\begin{tablenotes}
\item[a]\textit{Note:}The objective function was numerically maximized using differential evolution (DE) algorithm in \texttt{BlackBoxOptim.jl} package. For DE algorithm, I require setting the domain of parameters and the number of population seeds so that I fix the former to $[-10, 10]$ for parameters and the latter to 400. The reported point estimates are the best-found maxima. I omit the parameter $\beta_0$ that can take on only a finite number of values (here +1) converge at an arbitrarily fast rate, they are superconsistent. 
\end{tablenotes}
}\end{center}
\label{tb:estimation_results_single_market_two_param_beta_with_dummy_with_IR_penalty_20}
\end{table}

\end{document}